\documentclass[12pt]{article}

\usepackage{amssymb,amsmath,amsbsy,amsthm}
\usepackage{latexsym,graphicx}
\usepackage{authblk}

\newcommand{\newsection}{\setcounter{equation}{0}\section}

\renewcommand{\appendix}{\setcounter{equation}{0}\setcounter{section}{0}\renewcommand{\thesection}{\Alph{section}}}

\newcommand{\cc}{\hat\rho} 

\newtheorem{theorem}{Theorem}[section]
\newtheorem{lemma}[theorem]{Lemma}
\newtheorem{prop}[theorem]{Proposition}
\newtheorem{cor}[theorem]{Corollary}
\newtheorem{remark}[theorem]{Remark}

\newcommand{\pdag}{^{\phantom +}}
\newcommand{\Db}{\mathcal{D}_b}
\newcommand{\QED}{\hfill$\square$}

\newcommand{\ee}{{\rm e}}
\newcommand{\ii}{{\rm i}}
\newcommand{\half}{\mbox{$\frac{1}{2}$}}

\newcommand{\R}{{\mathbb R}}
\newcommand{\C}{{\mathbb C}}
\newcommand{\Z}{{\mathbb Z}}
\newcommand{\N}{{\mathbb N}}

\newcommand{\cF}{{\mathcal  F}}

\newcommand{\cH}{{\mathcal  H}}

\newcommand{\cN}{{\mathcal  N}}
\newcommand{\cQ}{{\mathcal  Q}}
\newcommand{\cC}{{\mathcal  C}}

\newcommand{\xxa}{\stackrel {\scriptscriptstyle \times}{\scriptscriptstyle \times} \!}
\newcommand{\xxe}{\! \stackrel {\scriptscriptstyle \times}{\scriptscriptstyle \times}}

\newcommand{\vzero}{\mathbf{0}}
\newcommand{\ve}{\mathbf{e}}
\newcommand{\vx}{\mathbf{x}}
\newcommand{\vy}{\mathbf{y}}
\newcommand{\vz}{\mathbf{z}}
\newcommand{\vw}{\mathbf{w}}

\newcommand{\vk}{\mathbf{k}}
\newcommand{\vlam}{\boldsymbol{\lambda}}
\newcommand{\vn}{\mathbf{n}}
\newcommand{\vm}{\mathbf{m}}
\newcommand{\vmu}{\boldsymbol{\mu}}
\newcommand{\veps}{\boldsymbol{\epsilon}}

\author[1,*]{Farrokh Atai}
\author[1,\dag]{Edwin Langmann}

\affil[1]{Department of Theoretical Physics, KTH Royal Institute of Technology \newline SE-106 91 Stockholm, Sweden \vspace{2mm}}

\title{\Large{\bf Deformed Calogero-Sutherland model and fractional Quantum Hall effect}} 

\date{\vspace{-1.0cm}\small\today}

\begin{document}

\maketitle

\let\oldthefootnote\thefootnote
\renewcommand{\thefootnote}{\fnsymbol{footnote}}
\footnotetext[1]{Electronic address: {\tt farrokh@kth.se}}
\footnotetext[2]{Electronic address: {\tt langmann@kth.se}}
\let\thefootnote\oldthefootnote

\vspace{-1.3cm}

\begin{abstract}
The deformed Calogero-Sutherland (CS) model is a quantum integrable system with arbitrary numbers of two types of particles and reducing to the standard CS model in special cases. 
We show that a known collective field description of the CS model, which is based on conformal field theory (CFT), is actually a collective field description of the deformed CS model. 
This provides a natural application of the deformed CS model in Wen's effective field theory of the fractional quantum Hall effect (FQHE), with the two kinds of particles corresponding to electrons and quasi-hole excitations.
In particular, we use known mathematical results about super Jack polynomials to obtain simple explicit formulas for the orthonormal CFT basis proposed by van Elburg and Schoutens in the context of the FQHE. 
\end{abstract}

\newsection{Introduction}
Calogero-Moser-Sutherland models have become a paradigm for exactly solvable systems which not only have fascinating mathematical properties but also interesting physics applications. 
In this paper we consider the quantum version of the $A_{N-1}$ trigonometric Calogero-Moser-Sutherland model known as {\em CS model}  \cite{C,Su}.

As discovered by Chalykh, Feigin, Veselov \cite{CFV} and Sergeev \cite{Sergeev}, the CS model allows for a generalization that is most natural from a mathematics point of view. 
This {\em deformed CS model} is quantum integrable \cite{Kh}, and it is related to Lie superalgebras \cite{SV} similarly as the standard CS model is related to Lie algebras \cite{OP2}. 
It was found that the {\em super Jack polynomials}, which were proposed as a natural mathematical generalization of the Jack polynomials \cite{KOO}, provide eigenfunctions of the deformed CS model in a natural generalization of a well-known result for the standard CS model \cite{Sergeev,SV2}. 
Moreover, intriguing relations between eigenfunctions of the CS model have been found by extending the mathematical theory of Jack polynomials \cite{Stanley,MacD} to the super case \cite{SV2,HL}. 
However, different from the standard CS model, no satisfactory interpretation of the deformed CS model as quantum many-body system is known. 
In this paper we present a generalization of a well-known collective field representation of the standard CS model \cite{MP,Iso,AMOS,CL} to the deformed case. 
As will be discussed, this provides a physics application of the deformed CS model to the fractional quantum Hall effect (FQHE) \cite{TSG}.
We also mention applications to the theory of special functions. 

We consider vertex operators $\phi_\nu(x)= \, \xxa \exp(-\ii\nu\varphi(x))\xxe$ in a standard conformal field theory (CFT) of chiral bosons $\varphi(x)$ such that $J(x)\equiv \partial_x \varphi(x)$ satisfies the affine Kac-Moody algebra relations $[J(x),J(y)] = -2\pi\ii\partial_x\delta(x-y)$, with $x,y$ variables on the circle $[-\pi,\pi)$ and $\nu$ a non-zero parameter such that $\nu^2$ is rational (see e.g.\ \cite{Kac}; precise mathematical definitions can be found also in the main text). 
This CFT allows for a {\em collective field description} \cite{JS} of the CS model. This provides a self-adjoint CFT operator $\cH^{\nu,3}$ that has the following relations with a product of an arbitrary number $N$ of such vertex operators, 
\begin{equation} 
\label{CSrel}
\cH^{\nu,3} \phi_\nu(x_1)\cdots \phi_\nu(x_N)|0\rangle  = H_{N}(\vx;\nu^2) \phi_\nu(x_1)\cdots \phi_\nu(x_N)|0\rangle 
\end{equation} 
with 
\begin{equation} 
\label{HNintro} 
H_{N}(\vx;g) = -\sum_{j=1}^N\frac{\partial^2}{\partial x_j^2} + \sum_{j<k}^N\frac{g (g -1)}{2\sin^2\frac12(x_j-x_{k})} 
\end{equation} 
the CS Hamiltonian with coupling parameter $g=\nu^2$ and $|0\rangle$ the vacuum \cite{CL}.  
This collective field description has many interesting implications explored in the physics and mathematics literature; see e.g.\ \cite{AMOS2,L1,SSAFR,SV3,EPSS,Nakajima}. 
We obtain a generalization of this to a product of arbitrary numbers $N$,$M$ of two types of vertex operators $\phi_\nu(x)$ and $\phi_{-1/\nu}(x)$ as follows, 
\begin{multline} 
\label{CSrel1}
\cH^{\nu,3} \phi_\nu(x_1)\cdots \phi_\nu(x_N)\phi_{-1/\nu}(y_1)\cdots \phi_{-1/\nu}(y_M)|0\rangle  = \\ 
\left( H_{N,M}(\vx,\vy;\nu^2) + \frac{1}{12}(\nu^2-\nu^{-2})M \right) \phi_\nu(x_1)\cdots \phi_\nu(x_N)\phi_{-1/\nu}(y_1)\cdots \phi_{-1/\nu}(y_M)|0\rangle 
\end{multline} 
with 
\begin{equation} 
\label{HNMintro} 
 H_{N,M}(\vx,\vy;g) = H_{N}(\vx;g) - g H_{M}(\vy;1/g) + \sum_{j=1}^N\sum_{k=1}^M \frac{(1-g )}{2\sin^2\frac12(x_j-y_k)}
\end{equation} 
the differential operator defining the deformed CS model \cite{Sergeev}; see Theorem~\ref{Thm:Main}  for a precise formulation. 
 It is important to note that the CFT operator $\cH^{\nu,3}$ appearing in \eqref{CSrel} and \eqref{CSrel1} is the same. 
 Thus our main results can be summarized as follows:  {\em The known collective field theory description of the CS model is, in fact, a  collective field theory description of the deformed CS model.}
This generalization is important in order to construct a complete basis of states in the pertinent Fock space: using \eqref{CSrel}, one can only generate a small subspace of states in this Hilbert space, and one needs \eqref{CSrel1} to get a complete basis 
 
It is worth noting that, in many works, the collective field theory description of the CS model is based on the space $\Lambda$ of symmetric functions in infinitely many variables $\vz=(z_1,z_2,\ldots)$, whereas we use a similar construction on a fermion Fock space $\cF$. These two approaches are related in that both are based on representations of the Heisenberg algebra, but there is one important difference: we use an extended Heisenberg algebra with a charge operator that has eigenvalues $Q\in\mathbb{Z}$, whereas the standard approach on the space of symmetric functions is restricted to $Q=0$ (see e.g.\ Equation (3.7) in \cite{Nakajima}, which corresponds to our Equation \eqref{cHnu3} for $\cQ=0$). It is possible to extend the latter approach to also allow for a charge operator, but then one has to work with the space $\tilde\Lambda=\bigoplus_{Q\in\mathbb{Z}}\Lambda_Q$ with infinitely many distinguished copies $\Lambda_Q$ of the space $\Lambda$. It is only in this setting that it becomes natural to extend previously known results about the CS model to the deformed CS model. Our motivation is the FQHE and, from this point of view, it is natural to start with the extended Heisenberg algebra. 
 
Our main application of \eqref{CSrel1} is in the context of Wen's theory of the FQHE \cite{Wen}, which is based on the very same CFT model that underlies our construction. 
In Wen's theory, the vertex operators $\phi_\nu(x)$ and $\phi_{-1/\nu}(x)$ play a fundamental role since they describe quasi-particle excitations \cite{ES}. 
However, since these quasi-particles are strongly interacting, the {\em anyon states}\footnote{We use this names since, in general, the vertex operators $\phi_{\nu}(x)$ and $\phi_{-1/\nu}(x)$ obey the exchange relations of anyons; see \eqref{exc}.} obtained by acting with the Fourier modes of these vertex operators on the vacuum $|0\rangle$ are not orthogonal, and this is a problem in practical computations. 
As proposed in \cite{ES}, it is possible to use the relation of this theory to the CS model to orthogonalize the anyon states. 
We use the result in \eqref{CSrel1} to construct an explicit map from super Jack polynomials to such {\em orthogonalized anyon states}; see Proposition~\ref{Prop:sJack}. 
The states obtained from Jack polynomials correspond to the special cases $(N,M)=(N,0)$ and $(0,M)$ but, to get a complete set of states, one needs super Jack polynomials allowing for arbitrary particle numbers $(N,M)$. 
As we discuss, results in the literature \cite{EPSS,ES,NS} suggest that the CFT states thus obtained form a complete orthogonal basis of common eigenstates of an infinite set of operators $\cH^{\nu,k}$, $k=1,2,\ldots $,  which should exist in this theory by general symmetry considerations \cite{Winfty}. 
In this paper we prove the common eigenstate property only for $k=1,2,3$, with $\cH^{\nu,1}$, $\cH^{\nu,2}$ and $\cH^{\nu,3}$ corresponding to the {\em charge operator}, the {\em CFT Hamiltonian}, and the {\em dCS collective field Hamiltonian}, respectively. 
To set this result in perspective, and to connect with previous results in the literature \cite{CL,ES}, we present a complimentary construction of such eigenstates in Appendix~\ref{sec:cHnu3}.

We also mention applications of our result in \eqref{CSrel1} to the theory of super Jack polynomials. 
In particular, we construct a map from eigenstates of the operators $\cH^{\nu,3}$ to eigenfunctions of the deformed CS model; see Corollary~\ref{cor:A}.  
Using this map and our construction of eigenstates of $\cH^{\nu,3}$ described above, we obtain an explicit integral transform between super Jack polynomials with different variable numbers $(N,M)$ and $(N',M')$; see \eqref{hPi}--\eqref{Pi1}. 
This also gives a quantum field theory interpretation of result previously obtained by other methods in \cite{HL}. 
We also mention that this map opens a way to generalize results in \cite{SV2}, but we leave a further investigation of this to future work. 

The relation between the CS model and the FQHE is known since a long time \cite{AI} and has been explored in many papers; see e.g.\ \cite{EPSS,BH} for recent work. 
Our result suggest that the CS model is only part of this story, and the deformed CS model is a natural generalization which can give a deeper understanding. 
It thus is natural to expect that other known results about the FQHE based on the standard CS model have interesting generalizations to the deformed case. 

It is interesting to note that the result in \eqref{CSrel1} was already anticipated in the physics literature on the FQHE: 
We mentioned already work of van Elburg and Schoutens \cite{ES} suggesting that the operator $\cH^{\nu,3}$ can be used to orthogonalize the anyon states. 
More specifically, they pointed out that the vertex operators $\phi_\nu(x)$ can only account for electron excitations, and to get a complete basis in the CFT Hilbert space one also needs to include the quasi-hole operators $\phi_{-1/\nu}(x)$. As an example they constructed an eigenstate of $\cH^{\nu,3}$ which contains one fermion and one quasi-hole and that corresponds to the case $N=M=1$ in our notation. 
We also mention that  mathematical details in the computations in \cite{LPS} could have been taken as a hint at the existence of super Jack polynomials,\footnote{We thank Didina Serban for discussions on this.} 
and that the parent Hamiltonian for particles and quasi-particles of the CS model proposed in \cite{AJ} is identical with the rational limit of the deformed CS model. 

The plan of this paper is as follows. The next section collects some definitions and know results about the deformed CS model. 
A precise formulation of our main result and its proof can be found in Section~\ref{sec:Main}. 
Our applications to the FQHE and special function theory are given in Sections~\ref{sec:sJack} and \ref{sec:Math}, respectively. 
Section~\ref{sec:Con} contains our conclusions. We also include two appendices. 
In Appendix~\ref{app:Proofs} we collect computational details and technical proofs. 
The alternative construction of eigenstates of dCS collective field Hamiltonian $\cH^{\nu,3}$ mentioned above can be found in Appendix~\ref{sec:cHnu3}.

A technical remark is in order: the framework we use is based on CFT results which are well-known in the physics literature. 
The present paper generalizes results in \cite{CL} where an analytic approach to making these CFT results precise is used. 
A more common such approach is based on the theory of vertex operators (see e.g.\ \cite{Kac} and references therein) which is somewhat closer to what is used in the physics literature than the approach in \cite{CL}. For the convenience of the reader we present our results using the latter approach but, for some technical proofs, we use results in \cite{CL}. 
More details on the relations between the two approaches, and a complementary derivation of the identity in \eqref{CSrel1}, can be found in \cite{FA}. 

\noindent {\bf Notation:} We use the symbol ``$\equiv$" to emphasize a definition. We denote as $\C$, $\R$, $\Z$ the complex, real, and integer numbers, $\N$ are the positive integers, and $\N_0\equiv\N\cup\{0\}$.  
 Moreover, $\Re(c)$ and $\Im(c)$ are the real and imaginary parts of a complex number $c$, respectively. 
 We do not distinguish a complex number $c$ and $c\times$(identity operator). 
We use physics notation in our construction of the CFT Hilbert space $\cF$; for example, $A^\dag$ is the Hilbert space adjoint of an operator $A$ on $\cF$, $\langle \eta |\eta'\rangle$ is the Hilbert space inner product of two states $\eta,\eta'\in\cF$, $|0\rangle$ is the vacuum in $\cF$, etc.; note also that $\langle\cdot|\cdot\rangle$ is linear in the second argument.  
 For vectors $\vx=(x_1,\ldots,x_N)$, we write $|\vx|$ short for $x_1+\cdots+x_N$. 
 Moreover, $\sum_{j<k}^N$ is short for $\sum_{1\leq j<k\leq N}$, etc.  
 We use standard definitions, terminology and results concerning partitions (see e.g.\  Macdonald's book \cite{MacD}, Section~I). 
 
 \newsection{Deformed Calogero-Sutherland model}
\label{sec:dCS} 
We collect definitions and mathematical results about the deformed CS model that we need. 

\subsection{Definition and basic facts}
\label{subsec:basicfacts} 
The deformed CS model is defined by the differential operator
\begin{equation} 
\label{HNM} 
\begin{split} 
H_{N,M}(\vx,\vy;g)  = -\sum_{j=1}^N\frac{\partial^2}{\partial x_j^2} + \sum_{j<j'}^N\frac{g (g -1)}{2\sin^2\frac12(x_j-x_{j'})} \\ 
+g \sum_{k=1}^{M}\frac{\partial^2}{\partial y_k^2} + \sum_{k<k'}^M\frac{(g -1)/g }{2\sin^2\frac12(y_k-y_{k'})}\\  
+ \sum_{j=1}^N\sum_{k=1}^M \frac{(1-g )}{2\sin^2\frac12(x_j-y_k)}
\end{split} 
\end{equation} 
with $N,M$ non-negative integers, $x_j$ and $y_k$ variables in the interval $[-\pi,\pi]$, and $g $ a real coupling parameter (we write $\vx$ short for $(x_1,\ldots,x_N)$ etc.; note that \eqref{HNM} is equivalent to the formulas in \eqref{HNintro} and \eqref{HNMintro}). This differential operator has exact eigenfunctions of the form 
\begin{equation} 
\label{Psilambda}
\Psi_{\vlam}(\vx,\vy;g) = \Psi_0(\vx,\vy;g)P^{N,M}_{\vlam}(\vz,\vw;g)
\end{equation} 
with the following generalization of the CS model ground state 
\begin{equation} 
\label{Psi0}
\Psi_{0}(\vx,\vy;g) = \frac{\Bigl(\prod_{j<j'}^N\sin\frac12(x_j-x_{j'})\Bigr)^{g }\Bigl(\prod_{k<k'}^M\sin\frac12(y_k-y_{k'})\Bigr)^{1/g }}{\prod_{j=1}^N\prod_{k=1}^M\sin\frac12(x_j-y_k)}
\end{equation} 
and $P\equiv P^{N,M}_{\vlam}$ polynomials in the variables $z_j\equiv \exp(\ii x_j)$, $w_k\equiv \exp(\ii y_k)$ known as {\em super Jack polynomials} \cite{Sergeev,SV2}.
The super Jack polynomials are symmetric in the variables $\vz$ and $\vw$ separately, satisfy the conditions 
\begin{equation} 
\label{condition}
\left.\left(\frac{\partial}{\partial z_j} + g \frac{\partial}{\partial w_k}\right) P(\vz,\vw;g)\right|_{z_j=w_k}=0\quad \forall j,k, 
\end{equation} 
and are labeled by partitions $\vlam$ in the fat $(N,M)$-hook, i.e.\ $\vlam=(\lambda_1,\lambda_2,\ldots)$ with non-negative integers $\lambda_j$ such that $\lambda_1\geq \lambda_2\geq \ldots$ and $\lambda_{N+1}\leq M$. The differential operator in \eqref{HNM} defines a deformation of the standard CS model by the parameter $M$ in the sense of Chalykh, Feigin, and Veselov \cite{CFV}. 
This deformed CS model is quantum integrable, i.e., there exist (sufficiently many) differential operators commuting with $H_{N,M}$ in \eqref{HNM}, including the first order differential operator 
\begin{equation} 
\label{PNM} 
D_{N,M}(\vx,\vy) \equiv  -\ii \left( \sum_{j=1}^N \frac{\partial}{\partial x_j} +  \sum_{k=1}^M \frac{\partial}{\partial y_{k} }  \right), 
\end{equation} 
and the functions $\Psi^{N,M}_{\vlam}$ above are common eigenfunctions of all these differential operators with explicitly known eigenvalues. 
For example, the eigenvalues corresponding to $D_{N,M}$ and $H_{N,M}$ are 
\begin{subequations} 
\begin{equation} 
\label{E2lam}
E^{(2)}_{\vlam} =  \sum_{j} \lambda_j
\end{equation} 
and
\begin{equation} 
\label{E3lam} 
E^{(3)}_{\vlam} =  \sum_j\left[ \lambda^2_j - g(2j-1)\lambda_j  \right] + \frac{1}{12}g^2\left[ (N-M/g)^3-(N-M/g^3) \right] , 
\end{equation} 
\end{subequations} 
respectively (this follows from results in \cite{Sergeev,SV2}; see Appendix~\ref{app:vlam=0} for further details).

It is tempting to interpret the deformed CS model as a quantum many-body system describing two different kinds of identical particles, in generalization of an important physics interpretation of the standard CS model. 
However, this interpretation is problematic since the eigenstates in \eqref{Psilambda}--\eqref{Psi0} are not square-integrable in the usual sense and, at least for $g>0$, one of the particle mass parameters is negative. 
Moreover, by introducing Planck's constant $\hbar$, one can see that the classical limit $\hbar\to 0$ of the deformed CS model is independent of $M$, i.e., the deformed model does not have a classical counterpart \cite{CFV}. 

In the rest of this paper we sometimes abuse physics terminology and refer to $H_{N,M}$ in \eqref{HNM} and $\Psi_0$ in \eqref{Psi0} as {\em dCS Hamiltonian} and {\em dCS groundstate}, respectively. 
 
\subsection{Super Jack polynomials and pseudo-momenta}
\label{subsec:superJack} 
We give further details about the super Jack polynomials. 
We also introduce finite integer vectors $\vn$ that are in one-to-one correspondence with partitions $\vlam$ in the fat $(N,M)$-hook and that we need later on. 

Let $\Lambda$ be the algebra of symmetric functions in infinitely many variables $\vz=(z_1,z_2,\ldots)$, $p_n(\vz)=z^n_1+z^n_2+\ldots$ the power sums ($n\in\mathbb{N}$), and $P^{(1/g)}_{\vlam}(\vz)$ the Jack polynomials with $\vlam=(\lambda_1,\lambda_2,\ldots)$ a partition; see e.g.\ Section~VI in \cite{MacD} (note that $g$ is equal to the inverse of the Macdonald parameter $\alpha$ \cite{MacD} and equal to the parameters $k$ and $\theta$ in \cite{Sergeev} and \cite{SV2}, respectively). The Jack polynomials have an expansion 
\begin{equation} 
\label{Jack} 
P^{(1/g)}_{\vlam}(\vz)= \sum_{\vmu} c^{(1/g)}_{\vlam,\vmu} p_{\vmu}(\vz),\quad p_{\vmu} = p_{\mu_1}p_{\mu_2}\cdots
\end{equation} 
with known coefficients $c^{(1/g)}_{\vlam,\vmu}$ (they are denoted as $\theta^{\;\;\vlam}_{\vmu}(1/g)/c_{\vlam}(1/g)$ in Macdonald's book \cite{MacD}; see Eqs.\ (10.22) and (10.28) in Section~VI in {\em loc.\ cit.}). 

Let $\Lambda_{N,M,g}$ be the algebra of polynomials in $(\vz,\vw)  =  (z_1,\ldots,z_N,  w_1,\ldots,w_M)$ which are symmetric in the variables $\vz$ and $\vw$ separately and which satisfy the condition in \eqref{condition}. The {\em deformed Newton sums}
\begin{equation} 
\label{spNMg} 
p^{N,M}_{n}(\vz,\vw;g)= \sum_{j=1}^N z_j^n - \frac1{g}\sum_{k=1}^M w_k^n 
\end{equation} 
obviously belong to $\Lambda_{N,M,g}$ for all non-negative integers $n$. The super Jack polynomials are defined as \cite{KOO} 
\begin{equation} 
\label{sJack} 
P^{N,M}_{\vlam}(\vz,\vw;g) = \sum_{\vmu} c^{(1/g)}_{\vlam,\vmu} p^{N,M}_{\vmu}(\vz,\vw;g),\quad p^{N,M}_{\vmu} = p^{N,M}_{\mu_1}p^{N,M}_{\mu_2}\cdots
\end{equation} 
with the very same coefficients $c^{(1/g)}_{\vlam,\vmu}$ as in \eqref{Jack}. 
Thus \eqref{Jack}--\eqref{sJack} provide explicit formulas for the super Jack polynomials; a more explicit formula can be found in \cite{SV2}, Eq.\ (25). 

As mentioned above, the partitions $\vlam$ labeling super Jack polynomials $P^{N,M}_{\vlam}$ are constrained by the condition $\lambda_{N+1}\leq M$. 
It is easy to see that there is a bijection between such partitions and integer vectors $\vn=(n_1,n_2,\ldots,n_{N+M})\in\mathbb{N}_0^{N+M}$ satisfying
\begin{equation} 
\label{nQuantumNrs} 
n_1\geq \cdots \geq n_N\geq K,\quad n_{N+1}\geq \cdots \geq n_{N+K}> n_{N+K+1}=\cdots=n_{N+M}=0 
\end{equation} 
for some $K$ in the range $0\leq K\leq M$,\footnote{Note that, for $K=0$, $n_j=0$ $\forall j>N$.} as follows, 
\begin{equation} 
\label{bijection}
n_j = \begin{cases} \lambda_j & (1\leq j\leq N)\\ \mu_{j-N}' & (N+1\leq j\leq N+M) \end{cases} 
\end{equation} 
with $\vmu=(\mu_1,\ldots,\mu_{M})$ the partition of length $M$ given by $\mu_j=\lambda_{N+j}$ and $\vmu'$ the partition conjugate to $\vmu$ (this works since $K\equiv \lambda_{N+1}$ is equal to the number of non-zero elements in $\vmu'$, and $K\leq M$) \cite{HL}.
We now define common eigenfunctions of these differential operators which generalize those in \eqref{Psilambda} and where we introduce a notation which is convenient later on, 
\begin{equation} 
\label{Psin} 
\Psi^{N,M}_{q_0,\vn}(\vx,\vy;g) = \ee^{\ii q_0(|\vx|-|\vy|/g)} \Psi^{N,M}_0(\vx,\vy;g)P^{N,M}_{\vlam}(\vz,\vw;g)  \quad (z_j=\ee^{\ii x_j}, w_k=\ee^{\ii y_k}) 
\end{equation} 
depending on an additional parameter $q_0$, and where $\vn\in\mathbb{N}_0^{N+M}$ is the integer vector determined by the partition $\vlam$ in the fat $(N,M)$-hook as in in \eqref{bijection} {\em ff.} (recall that $|\vx|$ is short for $x_1+\cdots+x_N$ etc.). 
We also find it convenient to re-define the function in \eqref{Psi0} as follows  
\begin{equation} 
\label{Psi01}
\Psi^{N,M}_{0}(\vx,\vy;g) = \frac{\Bigl(\prod^N_{j<j'}
\left[ \ee^{\ii (x_{j}-x_{j'})/2} - \ee^{\ii(x_{j'}-x_{j})/2}\right] 
\Bigr)^{g }\Bigl(\prod^M_{k<k'}
\left[ \ee^{\ii (y_{k}-y_{k'})/2} - \ee^{\ii(y_{k'}-y_{k})/2}\right] 
\Bigr)^{1/g }}{\prod_{j=1}^N\prod_{k=1}^M  
\left[ \ee^{\ii (x_{j}-y_{k})/2} - \ee^{\ii(y_{k}-x_{j})/2}\right] 
}
\end{equation} 
(note that this function is obtained from the one in \eqref{Psi0} by replacing $\sin(z)$ by $(\ee^{\ii z}-\ee^{-\ii z})$, i.e., both functions are equal up to an irrelevant constant).   
We note that, in the special case $M=0$, the exponential factor in \eqref{Psin} describes a center-of-mass motion which is often omitted in the standard CS model. 
However, this  exponential factor will be important for us.

One reason why the integer vectors $\vn$ in \eqref{bijection} are useful to label eigenfunctions is that they allow to express the eigenvalues of the deformed CS model \eqref{E2lam}--\eqref{E3lam} in terms of the following {\em pseudo-momenta} 
\begin{equation} 
\label{njplus}
n^+_j \equiv   \begin{cases} n_j + q_0 +\half g (N+1-2j)- \half M & (1\leq j\leq N) \\  n_j - g^{-1} q_0 +\half g^{-1}(M+1-2(j-N))+\half N & (1\leq j-N\leq M)  \end{cases} 
\end{equation} 
as follows \cite{HL}, 
\begin{subequations} 
\label{Ejn}
\begin{align} 
E^{(2)}_{q_0,\vn} & = \sum_{j=1}^{N+M} n_j^+, \label{E1n} \\
E^{(3)}_{q_0,\vn} & = \sum_{j=1}^N (n_j^+)^2 - g\sum_{j=1}^{M}(n_{N+j}^+)^2 \label{E2n}
\end{align} 
\end{subequations} 
(details on how \eqref{E1n} and \eqref{E2n} are obtained can be found in Appendix~\ref{app:vlam_vs_vn}).  

We conclude by writing down two identities for later reference, 
\begin{equation} 
\label{DPlamP}
\begin{split} 
D_{N,M}(\vx,\vy) \Psi^{N,M}_0(\vx,\vy;g) P^{N,M}_{\vlam}(\vx,\vy;g) &= |\vlam| \Psi^{N,M}_0(\vx,\vy;g) P^{N,M}_{\vlam}(\vx,\vy;g) ,\\
D_{N,M}(\vx,\vy) \Psi^{N,M}_0(\vx,\vy;g) p^{N,M}_{\vlam}(\vx,\vy;g) &= |\vlam| \Psi^{N,M}_0(\vx,\vy;g) p^{N,M}_{\vlam}(\vx,\vy;g) 
\end{split} 
\end{equation} 
(the first was already mentioned in words; the second is a simple consequence of definitions). 

\newsection{Main result} 
\label{sec:Main} 
We give the precise definition of the CFT we need and present our main result and its proof. The proof can be skipped without loss of continuity. 

\subsection{Prerequisites}
\label{subsec:CFT}
We consider a quantum field theory model defined by operators $\cc_n$, $n\in\mathbb{Z}$, and $R$ satisfying the relations
\begin{equation} 
\label{CCR} 
[\cc_n,\cc_m]=n\delta_{n+m,0},\quad [\cc_n,R] = \delta_{n,0}R,\quad \cc_n^\dag=\cc_{-n},\quad R^\dag = R^{-1} 
\end{equation} 
and which act on a Hilbert space $\mathcal{F}$ containing a vacuum state $|0\rangle$ such that 
\begin{equation} 
\label{HWcond}
\cc_n|0\rangle =0\quad \forall n\geq 0,\quad \langle 0|R^Q |0\rangle  = \delta_{Q,0}\quad \forall Q\in\mathbb{Z}  
\end{equation} 
(we set $R^0\equiv I$ and $R^{-Q}\equiv (R^\dag)^Q$). 
We recall the construction of $\cF$: the relations above imply that 
\begin{equation} 
\label{etab} 
\eta^b_{Q,\vm} \equiv \left( \prod_{n=1}^\infty  \frac{1}{\sqrt{n^{m_n}m_n!}}\cc_{-n}^{m_n}\right)R^Q|0\rangle \quad (m_n\in\mathbb{N}_0,\quad Q\in\mathbb{Z}), 
\end{equation} 
with only finitely many $m_n$ non-zero, are well-defined orthonormal states ($\vm$ is short for $(m_1,m_2,\ldots)$). 
We denote the set of {\em finite}  linear combinations of states in \eqref{etab} as $\Db$. 
This is obviously a pre-Hilbert space. The Hilbert space $\cF$ is obtained from $\Db$ by completion.

We note that $\cc_0$ has the physical interpretation as a charge operator,  and thus $R$ is a charge raising operator. 

We recall the definition of {\em normal ordering} which we denote by $\xxa\cdots\xxe$: it is defined by linearity and following rules, 
\begin{equation} 
\label{noord} 
\xxa R^Q\xxe \equiv R^Q,\quad \xxa M\cc_n\xxe = \xxa \cc_n M\xxe \equiv  \begin{cases} \xxa M\xxe\cc_n & \mbox{ if $n>0$}\\ (1/2)(\xxa M\xxe\cc_0+\cc_0\xxa M\xxe)& \mbox{ if $n=0$}\\  \cc_n\xxa M\xxe & \mbox{ if $n<0$}\end{cases} , 
\end{equation} 
for all integers $Q$ and operators $M$ which are products of powers of $R$ and an arbitrary finite number of operators $\cc_n$. 

We find it convenient to introduce a {\em statistics parameter unit} $\nu_0>0$ and the notation 
\begin{equation} 
\label{cQ} 
\cQ\equiv \nu_0\cc_0
\end{equation}  
The relations above imply $\cQ R = R(\cQ+\nu_0)$. 
Using this, vertex operators with statistics parameter $\nu\in\nu_0\mathbb{Z}$ can be defined as follows, 
\begin{equation} 
\label{phinu} 
\begin{split} 
\phi_\nu(x) & \equiv \,  \, \xxa R^{\nu/\nu_0} \exp\left( -\ii \nu \cQ x - \nu \sum_{n\neq 0} \frac{\cc_n}{n} \ee^{\ii nx}\right)\xxe\,  =  \\  
&  \exp{\left(-\nu \sum_{n<0}  \frac{\cc_n}{n}\ee^{\ii nx}\right)} \ee^{-\ii\nu \cQ x/2} R^{\nu/\nu_0} \ee^{-\ii\nu \cQ x/2} \exp{\left(-\nu\sum_{n>0} \frac{\cc_n}{n}\ee^{\ii nx}\right)} 
\end{split} 
\end{equation} 
with $x$ the position variable.    
The interpretation of these as quantum field operators on a circle suggests to restrict the $x$ to the interval $[-\pi,\pi]$, but, for technical reasons, it is convenient to analytically continue $x$ to the complex plane. 
It is worth mentioning the reason for our restriction on the statistics parameter: $\nu/\nu_0\in\mathbb{Z}$ is important since only integer powers of the operator $R$ are well-defined \cite{CR}. 

It is known that these vertex operators satisfy the exchange relations
\begin{equation}
\label{exc}  
\phi_\nu(x)\phi_{\nu'}(y) = \ee^{\mp \ii \pi \nu\nu'}\phi_{\nu'}(y)\phi_\nu(x) \text{ for }x\gtrless y; 
\end{equation} 
see e.g.\ \cite{CL}. In particular, if $\nu^2=m$ is an odd integer, then $\phi_{\nu}(x)$ are fermion operators; for $m=1$ they are standard chiral fermions, and for odd integers $m>1$ they correspond to {\em composite fermions} \cite{Wen}. 
However,  the results discussed here hold true for arbitrary real $\nu$ and, in general, the operators $\phi_\nu(x)$ correspond to {\em anyons} \cite{anyons}.    

Wen's effective theory of the FQHE at filling $1/m$, $m$ an odd integer, corresponds to the composite fermion case $\nu^2=m$ \cite{Wen}. 
As discussed in the next section, even if one is only interested in the composite fermions with statistics parameter $\nu=\sqrt{m}$, one is naturally led to also include anyons with statistics parameter $-1/\nu$. 

We conclude this section with a useful identity for a product of such anyon operators: 
\begin{equation} 
\label{phinuprod}
\phi_{\nu_1}(x_1) \cdots \phi_{\nu_K}(x_K) = \prod_{j<k}^K \left( \ee^{\ii(x_k-x_j)/2} -\ee^{\ii(x_j-x_k)/2}\right)^{\nu_j\nu_k}\xxa \phi_{\nu_1}(x_1)\cdots \phi_{\nu_K}(x_K) \xxa 
\end{equation} 
for all $K\in\mathbb{N}$, $\nu_j\in\nu_0\mathbb{Z}$, and $x_j\in\mathbb{C}$ such that  $0>\Im(x_1)>\cdots >\Im(x_K)$ (this follows from (70) and (72) in \cite{CL}, for example). 

\subsection{Collective field representation of the deformed CS model} 
\label{subsec:Main}
We consider operators $\mathcal{H}^{\nu,k}$ in this anyon model which implement non-negative integer powers of 
 the generator of translations as follows, 
\begin{equation} 
\label{cHnusdef}
\mathcal{H}^{\nu,k}\phi_\nu(x)|0\rangle  = \left(\ii \frac{\partial}{\partial x}\right)^{k-1}\phi_\nu(x)|0\rangle \quad (k=1,2,\ldots). 
\end{equation} 
For example, 
\begin{equation} 
\label{cHnu1} 
\mathcal{H}^{\nu,1}= \frac1{\nu}\cQ = \frac{\nu_0}{\nu}\cc_0
\end{equation} 
is the {\em charge operator}, and 
\begin{equation} 
\label{cHnu2} 
\mathcal{H}^{\nu,2} = \frac12 \cQ^2 + \sum_{n>0} \cc_{-n}\cc_n 
\end{equation} 
plays an important role as {\em CFT Hamiltonian}; see \cite{CL}. Our main object of interest is \cite{CL} 
\begin{equation}
\label{cHnu3}  
\cH^{\nu,3}= \frac{\nu}{3}\sum_{n,m,\ell\neq 0} \xxa \cc_{n}\cc_m\cc_\ell \xxe\delta_{\ell+m+n,0} +\sum_{n>0} [(1-\nu^2)n +2\nu \cQ]\cc_{-n}\cc_n +\frac{\nu}{3}\cQ^3-\frac{\nu^3}{12}\cQ . 
\end{equation} 

It is easy to see that the vacuum state $|0\rangle $ is a common eigenstate of the operators $\cH^{\nu,1}$, $\cH^{\nu,2}$, and $\cH^{\nu,3}$, and the corresponding eigenvalues are all 0. 
More generally, the states
\begin{equation} 
\label{OmQ} 
|Q\rangle \equiv R^Q|0\rangle \quad (Q\in\mathbb{Z}) 
\end{equation} 
are common eigenstates of these operators, and the corresponding eigenvalues are given by   
\begin{subequations} 
\label{cHnusOmega}
\begin{align} 
\mathcal{H}^{\nu,1} |Q\rangle & =  \frac{\nu_0Q}{\nu}|Q\rangle ,\label{cH1sOmega}\\
\mathcal{H}^{\nu,2} |Q\rangle & = \frac12(\nu_0 Q)^2|Q\rangle ,\label{cH2sOmega}\\
\mathcal{H}^{\nu,3} |Q\rangle & = \left( \frac{\nu(\nu_0Q)^3}3-\frac{\nu^3\nu_0Q}{12}\right) |Q\rangle .\label{cH3sOmega}
\end{align} 
\end{subequations} 

It is interesting to note the following simple  transformation properties of these operators under $\nu\to -1/\nu$: 
\begin{subequations} 
\begin{align} 
\mathcal{H}^{\nu,1} & = -\frac1{\nu^2}\mathcal{H}^{-1/\nu,1} \\
\mathcal{H}^{\nu,2} & = \mathcal{H}^{-1/\nu,2}\\ 
\mathcal{H}^{\nu,3} & = -\nu^2\mathcal{H}^{-1/\nu,3} +\frac1{12}(\nu^2 -\nu^{-2}) \mathcal{H}^{-1/\nu,1}.
\end{align} 
\end{subequations} 
We refer to these as {\em duality relations}� since, as shown in \cite{CL}, they are closely related to the celebrated duality of the Jack polynomials \cite{Stanley}. 
It is also worth mentioning that the following variants of these operators 
\begin{subequations} 
\begin{align} 
\tilde{\mathcal{H}}^{\nu,1} & = \nu\mathcal{H}^{\nu,1} = \cQ \\
\tilde{\mathcal{H}}^{\nu,2} & = \mathcal{H}^{\nu,2} = \frac12\cQ^2 + \sum_{n>0}\cc_{-n}\cc_n \\
\begin{split} 
\tilde{\mathcal{H}}^{\nu,3} & = \frac1{\nu}\mathcal{H}^{\nu,3}  +\frac1{12}\nu^2\cQ  \\ & = \frac{1}{3}
\mathop{\sum_{n,m,\ell\neq 0}}_{n+m+\ell=0}\xxa \cc_{n}\cc_m\cc_\ell \xxe  +\sum_{n>0} [(\nu^{-1}-\nu)n +2\cQ]\cc_{-n}\cc_n +\frac{1}{3}\cQ^3
\end{split} 
\end{align} 
\end{subequations} 
are invariant under this duality, i.e., 
\begin{equation} 
\tilde{\mathcal{H}}^{\nu,k} = \tilde{\mathcal{H}}^{-1/\nu,k}\quad (k=1,2,3).  
\end{equation} 

We now assume that also the operators $\phi_{-1/\nu}(x)$ are well-defined, i.e., not only $\nu$ but also $1/\nu$ is an integer multiple of $\nu_0$. 
It is easy to see that this imposes a restriction on $\nu_0$ and $\nu$: $\nu= r \nu_0$ and $1/\nu=s \nu_0$ imply $\nu_0=1/\sqrt{rs}$ and $\nu=\sqrt{r/s}$, or equivalently, 
 \begin{equation}
\label{rsdef}
\nu^2=\frac{r}{s},\quad \frac{\nu}{\nu_0}=r,\quad \frac1{\nu\nu_0}=s \quad (r,s\in\mathbb{N}).
\end{equation} 
It it natural to assume that $r$ and $s$ are co-prime (since then $r$ and $s$ are uniquely determined by $g=\nu^2$). 
However, the latter assumption is not necessary: many of our results hold true for an arbitrary pair of positive integers $(r,s)$ satisfying the condition $g=r/s>0$. This generalization is sometimes interesting; see, for example,  Lemma~\ref{lem:ES}.

The object of interest are the following products of anyons with statistics parameters $\nu$ and $-1/\nu$: 
\begin{equation} 
\label{PhinuNM} 
\Phi_\nu^{N,M}(\vx,\vy)\equiv\phi_\nu(x_1)\cdots \phi_\nu(x_N)\phi_{-1/\nu}(y_1)\cdots \phi_{-1/\nu}(y_{M}) . 
\end{equation} 
One can show that these are well-defined as quadratic forms provided that
\begin{equation} 
\label{Eq:restr} 
0>\Im(x_1)>\cdots >\Im(x_N)>\Im(y_1)>\cdots > \Im(y_M) 
\end{equation} 
(details to prove this can be found in \cite{LM}, Section~3.C, for example). 
Moreover, 
\begin{equation} 
\label{nphiNM}  
\Phi_\nu^{N,M}(\vx,\vy) =  \Psi^{N,M}_{0}(-\vx,-\vy;\nu^2)\xxa  \Phi_\nu^{N,M}(\vx,\vy) \xxe
\end{equation} 
with $\Psi^{N,M}_{0}$ the dCS groundstate in \eqref{Psi01} for $g=\nu^2$ (this is a special case of \eqref{phinuprod}). 
This identity allows to analytically continue the definition of the operators in \eqref{PhinuNM} to other values of $x_j$ and $y_{k}$.

We now are ready to state our result  (recall the definitions in \eqref{cHnu1}--\eqref{OmQ} and \eqref{PhinuNM}). 
  
\begin{theorem} 
\label{Thm:Main}  
The following hold true as identities of quadratic forms on $\Db$, 
\begin{subequations} 
\begin{align} 
\label{cH1nurel}
[\cH^{\nu,1},\Phi_{\nu}^{N,M}(\vx,\vy)]|Q\rangle  & = (N -M\nu^{-2}) \Phi_\nu^{N,M}(\vx,\vy)|Q\rangle , \phantom{\left(\frac12\right)}  \\
\label{cH2nurel} 
[\cH^{\nu,2},\Phi_{\nu}^{N,M}(\vx,\vy)]|Q\rangle  & =  - D_{N,M}(\vx,\vy) \Phi_\nu^{N,M}(\vx,\vy)|Q\rangle ,\phantom{\left(\frac12\right)} \\
\label{cH3nurel} 
[\cH^{\nu,3},\Phi_{\nu}^{N,M}(\vx,\vy)]|Q\rangle  & = \left( H_{N,M}(\vx,\vy;\nu^2) + \frac{1}{12}(\nu^2-\nu^{-2})M \right) \Phi_\nu^{N,M}(\vx,\vy)|Q\rangle , 
\end{align} 
with the differential operators $H_{N,M}$ and $D_{N,M}$ defined in \eqref{HNM} and \eqref{PNM}, respectively. 
\end{subequations} 
\end{theorem} 

(The proof is given in Section~\ref{subsec:Proof}.) 

As mentioned in the introduction, our main result is \eqref{cH3nurel}. 
However, we also need the identities in \eqref{cH1nurel} and \eqref{cH2nurel}. 
\subsection{Proof of Theorem~\ref{Thm:Main} }
\label{subsec:Proof} 
The proof is based on results obtained in \cite{CL}, with technical improvements in \cite{L,Lrem}.

Let
\begin{equation} 
\tilde\rho_\varepsilon(y)\equiv \frac1{2\pi}\left( \nu_0 \cc_0 + \sum_{n\neq 0} \cc_n\ee^{\ii ny-\varepsilon|n|} \right) \quad (\varepsilon>0,\; y\in [-\pi,\pi]). 
\end{equation} 
As explained in \cite{CL}, the operators
\begin{equation} 
\begin{split} 
W^{\nu,1} &= \left. \int_{-\pi}^\pi dy \xxa  \tilde\rho_\varepsilon(y)\xxe\right|_{\varepsilon=0}  \\
W^{\nu,2} &= \left.\pi\int_{-\pi}^\pi dy \xxa \tilde\rho_\varepsilon(y)^2\xxe\right|_{\varepsilon=0}  \\
W^{\nu,3} &= \left.\frac{4\pi^2}{3}\int_{-\pi}^\pi dy \xxa \tilde\rho_\varepsilon(y)^3\xxe\right|_{\varepsilon=0}  -\frac{\nu^2}{12}W^{\nu,1}  
\end{split} 
\end{equation} 
obey the identities
\begin{equation} 
\label{Wnu12rel}
[W^{\nu,1},\phi_\nu(x)]=\nu\phi_\nu(x),\quad 
[W^{\nu,2},\phi_\nu(x)] = \ii \frac{\partial}{\partial x}\phi_\nu(x)
\end{equation} 
and 
\begin{equation} 
\label{Wnu3rel}
[W^{\nu,3},\phi_\nu(x)] = -\frac1{\nu}\frac{\partial^2}{\partial x^2}\phi_\nu(x) + 2\pi\ii(\nu^2-1)\xxa \rho'(x) \phi_\nu(x)\xxe 
\end{equation} 
with 
\begin{equation} 
\rho'(x) \equiv \frac1{2\pi} \sum_{n\neq 0} \ii n \cc_n\ee^{\ii nx} 
\end{equation} 
provided that $\Im(x)<0$ (this follows from Eqs.\ (87), (96) and (97) in \cite{CL}; a typo in Eq.\ (87) in \cite{CL} is herewith corrected). 
Moreover, 
\begin{equation} 
\label{Wnus} 
\begin{split} 
W^{\nu,1} & = \cQ \\
W^{\nu,2} & = \frac12 \cQ^2 + \sum_{n>0} \cc_{-n}\cc_n  \\ 
W^{\nu,3} & = \frac13 \mathop{\sum_{n,m,\ell\neq 0}}_{n+m+\ell=0}\xxa \cc_{n}\cc_m\cc_\ell \xxe + 2\cQ\sum_{n>0}\cc_{-n}\cc_n + \frac13\cQ^3 -\frac{\nu^2}{12}\cQ
\end{split} 
\end{equation} 
(this is essentially Eq.\ (88) in \cite{CL}). From the formulas above it is obvious that 
\begin{equation} 
\label{WnusOmega}
W^{\nu,k}|0\rangle =0\quad (k=1,2,3).
\end{equation} 
This and \eqref{Wnu12rel} imply \eqref{cHnusdef} for $k=1,2$ with $\cH^{\nu,1}=W^{\nu,1}/\nu$, $\cH^{\nu,2}=W^{\nu,2}$. 
From this it is easy to obtain \eqref{cHnu1} and \eqref{cHnu2}.

For later use we note that, if $\Im(x)<0$, 
\begin{equation}
\label{Wnu12rel2}
[W^{\nu,1},\phi_{-1/\nu}(x)]=-\frac{1}{\nu}\phi_{-1/\nu}(x),\quad 
[W^{\nu,2},\phi_{-1/\nu}(x)] = \ii \frac{\partial}{\partial x}\phi_{-1/\nu}(x)
\end{equation} 
and 
\begin{equation} 
\label{Wnu3rel2} 
[W^{\nu,3},\phi_{-1/\nu}(x)] = 
\left(  \nu\frac{\partial^2}{\partial x^2} + \frac1{12}(\nu-\nu^{-3})\right) \phi_{-1/\nu}(x) +  2\pi\ii(\nu^{-2}-1)\xxa \rho'(x) \phi_{-1/\nu}(x)\xxe, 
\end{equation} 
which follows from the following relations implied by the explicit formulas for these operators, $W^{\nu,1}=W^{-1/\nu,1}$, $W^{\nu,2}=W^{-1/\nu,2}$ and $W^{\nu,3} = W^{-1/\nu,3} - \frac1{12}(\nu^2-\nu^{-2})W^{-1/\nu,1}$ (cf.\ \eqref{Wnus}). 

The definitions in \eqref{cHnu1} and \eqref{cHnu2} and the relations in \eqref{Wnu12rel} and \eqref{WnusOmega} trivially imply  \eqref{cHnusdef} for $k=1,2$. 
To prove  \eqref{cHnusdef} for $k=3$ we use the operator 
\begin{equation} 
\label{Cdef} 
\mathcal{C}\equiv  \sum_{n>0} n \cc_{-n}\cc_n 
\end{equation}  
which is known to satisfy the relations
\begin{equation} 
\label{Crel} 
\mathcal{C}\phi_\nu(x) + \phi_\nu(x)\mathcal{C} = 2\pi\ii\nu \xxa\rho'(x)\phi_\nu(x) \xxe + 2\xxa \mathcal{C} \phi_\nu(x)\xxe, \quad \mathcal{C}|0\rangle  = 0
\end{equation} 
provided that $\Im(x)<0$ (see Eqs.\ (99), (101) and (104) in \cite{CL}). 
This implies that the operator 
\begin{equation} 
\label{cHnu31}
\mathcal{H}^{\nu,3}\equiv \nu W^{\nu,3} + (1-\nu^2)\mathcal{C} , 
\end{equation} 
which is equal to the one in \eqref{cHnu3}, satisfies \eqref{cHnusdef} for $k=3$ (cf.\  Eq.\ (105) in \cite{CL}). 

We now are ready to prove  \eqref{cH1nurel}--\eqref{cH3nurel}. In the following we assume \eqref{Eq:restr}. We find it convenient to use the notation 
\begin{equation} 
\label{DefProof}
\begin{split} 
\Phi_\nu^{M,N}(\vx,\vy) = \phi_{\nu_1}(x_1) \cdots \phi_{\nu_{K}}(x_K) ,\quad K=N+M , \\ (\nu_j,x_j) = \begin{cases} (\nu,x_j) & \mbox{ for } j=1,\ldots,N, \\ (-1/\nu,y_{j-N}) & \mbox{ for } j=N+1,\ldots,N+M, \end{cases}
\end{split} 
\end{equation} 
to compute 
\begin{equation} 
[W^{\nu,1},\Phi_\nu^{M,N}(\vx,\vy)]  =    \sum_{j=1}^K \phi_{\nu_1}(x_1) \cdots [W^{\nu,1},\phi_{\nu_j}(x_j) ]\cdots \phi_{\nu_{K}}(x_K)  
=  \left(N\nu-M\frac{1}{\nu} \right) \Phi_\nu^{M,N}(\vx,\vy)  
\end{equation} 
and, similarly, 
\begin{equation} 
\begin{split} 
[W^{\nu,2},\Phi_\nu^{M,N}(\vx,\vy)]  =  
\ii \sum_{j=1}^K \frac{\partial}{\partial x_j}\Phi_\nu^{M,N}(\vx,\vy)  
\end{split} 
\end{equation} 
using \eqref{Wnu12rel} and \eqref{Wnu12rel2}, which imply the relations in  \eqref{cH1nurel} and \eqref{cH2nurel}. 
To prove  \eqref{cH3nurel} we compute, in a similar manner, 
\begin{multline} 
[W^{\nu,3},\Phi_\nu^{M,N}(\vx,\vy)]  =  \left( -\sum_{j=1}^K\frac1{\nu_j}\frac{\partial^2}{\partial x_j^2} + \frac1{12}(\nu-\nu^{-3}) M \right) \Phi_\nu^{M,N}(\vx,\vy)  + \\  
  \quad 2\pi\ii\sum_{j=1}^K(\nu_j^2-1)\phi_{\nu_1}(x_1)\cdots  \xxa \rho(x_j)' \phi_{\nu_j}(x_j) \xxe \cdots \phi_{\nu_{K}}(x_{K})  
\end{multline} 
using \eqref{Wnu3rel} and \eqref{Wnu3rel2}. 
To proceed we use the following generalization of \eqref{Crel} \cite{CL,L} 
\begin{multline} 
\mathcal{C} \Phi_\nu^{M,N}(\vx,\vy) +  \Phi_\nu^{M,N}(\vx,\vy) \mathcal{C}   = 2\, \mathcal{C}*\Phi_\nu^{M,N}(\vx,\vy) + \\ 
2\pi\ii \sum_{j=1}^K \nu_j \left( \rho'_+(x_j)\Phi_\nu^{M,N}(\vx,\vy) +  \Phi_\nu^{M,N}(\vx,\vy) \rho'_-(x_j) \right)   
\end{multline} 
with 
\begin{equation} 
\label{rhopm} 
\rho'_+(x) = \frac1{2\pi}\sum_{n<0} \ii n\cc_n \ee^{\ii nx} \; \mbox{ and }\; \rho'_-(x) = \frac1{2\pi}\sum_{n>0} \ii n\cc_n \ee^{\ii nx}
\end{equation} 
the creation- and annihilation part of $\rho'(x)$ and 
\begin{equation} 
\mathcal{C}*\Phi_\nu^{M,N}(\vx,\vy) \equiv \sum_{n>0} n\cc_{-n} \Phi_\nu^{M,N}(\vx,\vy) \cc_n 
\end{equation} 
(this is a simple generalization of Lemma~4 in \cite{CL} which can be obtained from Lemma~3 in  \cite{L} as a special case). 
Recalling \eqref{cHnu31}, and using that $(1-\nu^2)\nu_j = -\nu(\nu_j^2-1)$ for all $j$,  we obtain 
\begin{multline} 
\label{cH3nu10}
[\mathcal{H}^{\nu,3}, \Phi^{N,M}_\nu(\vx,\vy)]   = 2(1-\nu^2)\, \mathcal{C}*\Phi^{N,M}_\nu(\vx,\vy) - 2(1-\nu^2)\Phi^{N,M}_\nu(\vx,\vy)\mathcal{C} +  \\
 \left( -\sum_{j=1}^K\frac{\nu}{\nu_j}\frac{\partial^2}{\partial x_j^2} + \frac1{12}(\nu^2-\nu^{-2}) M \right) \Phi_\nu^{M,N}(\vx,\vy) + \sum_{j=1}^K 2\pi\ii\nu(\nu_j^2-1)\times \\
\Bigl( [\phi_{\nu_1}(x_1)\cdots \phi_{\nu_{j-1}}(x_{j-1}),\rho'_+(x_j)] \phi_{\nu_j}(x_j) \cdots \phi_{\nu_{K}}(x_{K}) + \\ \phi_{\nu_1}(x_1)\cdots \phi_{\nu_j}(x_j)[\rho'_-(x_j),\phi_{\nu_{j+1}}(x_{j+1}) \cdots \phi_{\nu_{K}}(x_{K})] \Bigr)
\end{multline} 
(we inserted $ \xxa \rho'(x_j) \phi_{\nu_j}(x_j) \xxe =  \rho'_+(x_j) \phi_{\nu_j}(x_j) + \phi_{\nu_j}(x_j) \rho'_-(x_j)$ which follows from the definition of normal ordering). 
We now use \cite{CL,L}
\begin{equation} 
\label{rhopmrels}
\begin{split} 
[\rho'_-(x_j),\phi_{\nu_k}(x_k)] &=  -\frac{\ii}{4\pi}\nu_k V(x_j-x_k) \phi_{\nu_k}(x_k) \quad (j<k),\\
[\phi_{\nu_k}(x_k),\rho'_+(x_j)] &= -\frac{\ii}{4\pi}\nu_k V(x_k-x_j) \phi_{\nu_k}(x_k) \quad (j>k) 
\end{split} 
\end{equation} 
with 
\begin{equation} 
\label{Vdef} 
V(x) \equiv -2\sum_{n>0} n\ee^{\ii n x} = \frac1{2\sin^2(\half x)}\quad (\Im(x)>0)
\end{equation} 
(for the convenience of the reader we give a proof of \eqref{rhopmrels} at the end of this section). 
This yields
\begin{multline} 
\label{cH3nu100}
[\mathcal{H}^{\nu,3},\Phi^{N,M}_\nu(\vx,\vy)]   =  2(1-\nu^2)\, \mathcal{C} *\Phi^{N,M}_\nu(\vx,\vy) - 2(1-\nu^2)\Phi^{N,M}_\nu(\vx,\vy)\mathcal{C} + \\
 \left( -\sum_{j=1}^K\frac{\nu}{\nu_j}\frac{\partial^2}{\partial x_j^2} + \frac1{12}(\nu^2-\nu^{-2}) M \right) \Phi_\nu^{M,N}(\vx,\vy) + \\
 \sum_{j<k}^K \nu\nu_k (\nu_j^2-1)V(x_j-x_k)  \Phi_\nu^{M,N}(\vx,\vy) .
\end{multline} 
Using 
\begin{equation} 
\label{gammajk} 
\gamma_{jk}\equiv \nu\nu_k(\nu_j^2-1) = \begin{cases} \nu^2(\nu^2-1) & \mbox{ if } j,k\leq N\\ (\nu^2-1)/\nu^2 & \mbox{ if } j,k>N\\ (1-\nu^2) & \mbox{ otherwise } \end{cases} 
\end{equation} 
and recalling the definition in \eqref{HNM}� we can write this as
\begin{multline} 
\label{cH3nurelgen}
[\mathcal{H}^{\nu,3},\Phi^{N,M}_\nu(\vx,\vy)]   =  2(1-\nu^2)\, \mathcal{C}*\Phi^{N,M}_\nu(\vx,\vy) - 2(1-\nu^2)\Phi^{N,M}_\nu(\vx,\vy)\mathcal{C} +  \\ 
\left(H_{N,M}(\vx,\vy;\nu^2) + \frac1{12}(\nu^2-\nu^{-2}) M  \right)\Phi^{N,M}_\nu(\vx,\vy). 
\end{multline} 
The first two terms on the r.h.s.\  give zero when applied to any state $|Q\rangle$ (this follows from $\mathcal{C}|Q\rangle=0$ and the definition of normal ordering). 
Thus \eqref{cH3nurelgen} implies  \eqref{cH3nurel}. 

To conclude give the computation proving \eqref{rhopmrels}.  
 Inserting the definitions in \eqref{phinu} and \eqref{rhopm} we compute, using $[A,\ee^B]=[A,B]\ee^{B}$ for operators $A,B$ commuting with $[A,B]$,  and,  assuming $\Im(x-y)>0$, 
\begin{multline} 
[\rho'_-(x),\phi_{\nu_k}(y)] = 
\left[\frac1{2\pi}\sum_{n>0} \ii n \cc_n \ee^{\ii nx}, \exp{\left(-\nu_k \sum_{m<0}  \frac{\cc_m}{m} \ee^{\ii my}\right)}\right]   \ee^{-\ii\nu_k \cQ y/2}
R^{\nu_k/\nu_0} \ee^{-\ii\nu_k \cQ y/2} \times \\ \exp{\left(-\nu_k\sum_{m>0} \frac{\cc_m}{m}\ee^{\ii my}\right)} = 
-\nu_k\frac{\ii}{2\pi}\sum_{n>0} n \left[ \cc_n \ee^{\ii nx}, \sum_{m<0} \frac{\cc_m}{m}\ee^{\ii my} \right] \phi_{\nu_k}(y) = \\  
\nu_k\frac{\ii}{2\pi} \sum_{n>0} n \ee^{\ii n(x-y)}\phi_{\nu_k}(y) = -\nu_k\frac{\ii}{4\pi}V(x-y), 
\end{multline} 
which gives the first identity in \eqref{rhopmrels}. The second identity is proved in a similar manner. 

\hfill \QED

\newsection{Diagonalization of the collective field Hamiltonian}
\label{sec:sJack}
We show that the collective field Hamiltonian $\cH^{\nu,3}$ of the deformed CS model in \eqref{cHnu3} defines an exactly solvable system in the sense that its eigenstates and corresponding eigenvalues can be constructed explicitly. 
Our result is based on Theorem~\ref{Thm:Main}  and the theory of super Jack polynomials \cite{Sergeev,SV2}; see Proposition~\ref{Prop:sJack}. 
 
As mentioned in the introduction, we give a complementary construction of the eigenstates of $\cH^{\nu,3}$ in Appendix~\ref{sec:cHnu3}. 

\subsection{Super Jack states} 
As discussed in Section~\ref{subsec:superJack}, from the mathematical theory of super Jack polynomials we know explicit formulas for common eigenfunctions $\Psi^{N,M}_{q_0,\vn}$,  and corresponding eigenvalues, of the deformed CS  differential operators $H_{N,M}$ in \eqref{HNM} and $D_{N,M}$ in \eqref{PNM}; see \eqref{Jack}--\eqref{E2n}. 
For such an eigenfunction, we construct a corresponding CFT state as follows, 
\begin{equation} 
\label{psiNMnQ} 
\psi^{N,M}_{Q,\vn} \equiv \int_C d^Nx\, d^My\, \Psi^{N,M}_{q_0,\vn}(\vx,\vy;\nu^2) \Phi^{N,M}_\nu(\vx,\vy) |Q\rangle  
\end{equation} 
where we ignore some technical details for now (the precise definition of the integration is in  \eqref{intcC}--\eqref{q0} below). 
It turns out that, for suitable $q_0$ and suitable integration contours,  the integral in \eqref{psiNMnQ} defines a state in $\Db$. 
Moreover, the following simple computation, which uses Theorem~\ref{Thm:Main}  and that $\Psi^{N,M}_{q_0,\vn}$ in \eqref{Psin} is an eigenstate of $H_{N,M}$ with known eigenvalue $E^{(3)}_{q_0,\vn}$ in \eqref{E2n}, suggests that such state is an exact eigenstates of $\cH^{\nu,3}$: 
\begin{multline} 
\label{cHnu3comp}
\cH^{\nu,3}\psi^{N,M}_{Q,\vn} = \int d^Nx\,d^My\, \Psi^{N,M}_{q_0,\vn}(\vx,\vy;\nu^2) \cH^{\nu,3} \Phi^{N,M}_\nu(\vx,\vy) |Q\rangle =\\
\int  d^Nx\,d^My\,\Psi^{N,M}_{q_0,\vn}(\vx,\vy;\nu^2) \left( H_{N,M}(\vx,\vy;\nu^2) +e_{3,Q} \right) \Phi^{N,M}_\nu(\vx,\vy) |Q\rangle =\\ 
\int  d^Nx\,d^My\,\left\{ \left( H_{N,M}(\vx,\vy;\nu^2) +e_{3,Q} \right) 
\Psi^{N,M}_{q_0,\vn}(\vx,\vy;\nu^2)\right\}    \, \Phi^{N,M}_\nu(\vx,\vy;\nu^2) |Q\rangle = \\
\left( E^{(3)}_{q_0,\vn}+ e_{3,Q} \right) \psi^{N,M}_{Q,\vn} 
\end{multline} 
with the constant $e_{3,Q}=\frac1{12}(\nu^2-\nu^{-2})M+\frac13\nu(\nu_0Q)^3-\frac1{12}\nu^3\nu_0Q$, where the last line gives the corresponding eigenvalues (we used \eqref{cH3sOmega} and \eqref{cH3nurel} in the first equality and made two partial integrations in the second).  
Corresponding results for the operators $\cH^{\nu,1}$ and $\cH^{\nu,2}$ are obtained in a similar manner. 

To make the formula in \eqref{psiNMnQ} precise we specify the integration domain as follows
\begin{equation} 
\label{intcC} 
\int_Cd^Nx\,d^My\,\equiv \int_{C_1} dx_1\cdots \int_{C_N}dx_N \int_{C_{N+1}} dy_1\cdots \int_{C_{N+M}}dy_M
\end{equation} 
with the integration contours\footnote{It follows from Cauchy's theorem that there is a large freedom to deform these integration contours.} 
\begin{equation} 
\label{cCeps} 
\begin{array}{lll} 
C_j:&\!\!\! x_j(t) =  t -\ii \epsilon_1 -\ii (\epsilon_j-\epsilon_1) \cos^2(t/2) &  (-\pi\leq t\leq \pi,\; j=1,\ldots,N)\\
C_{N+k}:&\!\!\! y_k(t) =  t -\ii \epsilon_{N+1} -\ii (\epsilon_{N+k}-\epsilon_{N+1}) \cos^2(t/2) &  (-\pi\leq t\leq \pi,\; k=1,\ldots,M)
\end{array} 
\end{equation}  
and parameters $\epsilon_j$ satisfying 
\begin{equation} 
\label{epsorder1} 
0<\epsilon_1<\epsilon_2<\cdots<\epsilon_{N+M}. 
\end{equation} 
We also recall the definitions in \eqref{etab} {\em ff.} and \eqref{PhinuNM}, and that $|\vx|$ is short for $\sum_j x_j$.

 \begin{prop} 
 \label{Prop:sJack} 
For $g=\nu^2>0$ rational,  $\Psi_0^{N,M}$ the dCS groundstate in \eqref{Psi01}, $P^{N,M}_{\vlam}$ the super Jack polynomial in \eqref{Jack}--\eqref{sJack}, $Q\in\mathbb{Z}$, and the integer vectors $\vn\in\mathbb{N}_0^{N+M}$ determined by $\vlam$ as in \eqref{bijection}, let 
\begin{equation} 
\label{psiNMnQ1} 
\psi^{N,M}_{Q,\vn} \equiv \int_Cd^Nx\,d^My\,\ee^{\ii q_0(|\vx|-|\vy|/\nu^2)}\Psi_0^{N,M}(\vx,\vy;\nu^2) P^{N,M}_{\vlam}(\vz,\vw;\nu^2)  \Phi^{N,M}_\nu(\vx,\vy) |Q\rangle 
\end{equation} 
with $z_j=\ee^{\ii x_j}$, $w_k=\ee^{\ii y_k}$, $\Phi_\nu^{N,M}(\vx,\vy)$ as in \eqref{nphiNM}, and the integration defined in \eqref{intcC}--\eqref{epsorder1}. 
Then this integral is finite provided 
\begin{equation}
\label{q0} 
q_0 = \half(N\nu^2-M) + \nu_0\nu Q, 
\end{equation}
and it defines a state in $\Db$ that is independent of $\veps$. 
Moreover, $\psi^{N,M}_{Q,\vn}$ is a common eigenstate of the operators $\cH^{\nu,1}$, $\cH^{\nu,2}$, $\cH^{\nu,3}$ in \eqref{cHnu1}--\eqref{cHnu3} with corresponding eigenvalues given by 
\begin{subequations} 
\label{EkNMnun} 
\label{eigenvalues} 
\begin{align} 
E^{N,M}_{1,Q} = & (N-\nu^{-2}M) +  \nu^{-1}\nu_0Q,\label{E1NMnun}\\
E^{N,M}_{2,Q,\vn}  = & \sum_{j=1}^{N+M} n^+_j + \half(\nu_0Q)^2,\label{E2NMnun}\\
E^{N,M}_{3,Q,\vn} = & \sum_{j=1}^{N} (n^+_j)^2 -\nu^2\sum_{j=1}^{M}(n^+_{N+j})^2 + \frac1{12}(\nu^2-\nu^{-2})M +\frac{\nu(\nu_0Q)^3}{3}-\frac{\nu^3\nu_0Q}{12}, \label{E3NMnun}
\end{align} 
\end{subequations} 
with 
\begin{equation} 
\label{pjdef} 
n^+_j =  \begin{cases} n_j + \nu^2(N+\half -j)-M+\nu\nu_0Q   & (1\leq j\leq N) , \\ 
n_j + \nu^{-2}(M+\half -(j-N)) -\nu^{-1}\nu_0 Q & (1\leq j-N\leq M). \end{cases}
\end{equation} 
 \end{prop} 

(A proof, which includes explicit formulas for the states $\psi^{N,M}_{Q,\vn}$, is given in  Section~\ref{subsec:ProofThm:sJack}.)
 
 We mention that integration contours as in \eqref{cCeps}--\eqref{epsorder1} are well-known in CFT \cite{Felder}.

Note that $n_j^+$ in \eqref{pjdef} is equal to the pseudo-momenta of the deformed CS model in \eqref{njplus} for $q_0$ in \eqref{q0} and $g=\nu^2$. 
Moreover, one can write the formulas for the eigenvalues in \eqref{EkNMnun} in a unified way as follows,
\begin{equation} 
\label{EkNMnun1} 
E^{N,M}_{k,Q,\vn} = \sum_{j=1}^N (n^+_j)^{k-1} + (-\nu^2)^{k-2}\sum_{j=1}^M(n^+_{N+j})^{k-1} + e_{k,Q}\quad (k=1,2,3)  
\end{equation} 
with constants $e_{k,Q}$ defined by this equation. 

It is interesting to note the following formula for the Hilbert space norm of the states in \eqref{psiNMnQ1} 
\begin{multline} 
\label{psinorm} 
||\psi^{N,M}_{Q,\vn}||^2 = \int_{\bar{C}} d^Nx\,d^My\,\int_C d^Nx'\, d^My'\, P^{N,M}_{\vlam}(\vz^{-1},\vw^{-1};\nu^2)\Delta^{N,M}(\vz^{-1},\vw^{-1};\nu^2) \times \\  
\Pi^{N,M}(\vz,\vw,(\vz')^{-1},(\vw')^{-1};\nu^2) \Delta^{N,M}(\vz',\vw';\nu^2)P^{N,M}_{\vlam}(\vz',\vw';\nu^2) , 
\end{multline} 
\begin{equation}
\label{Pi} 
\Pi^{N,M}(\vz,\vw,\vz',\vw';g) \equiv \frac{\prod_{j=1}^N\prod_{k=1}^M(1-z_jw_k')(1-z_j'w_k)}{\prod_{j,j'=1}^N(1-z_j z_{j'}')^{g}\prod_{k,k'=1}^M(1-z_kz_{k'}')^{1/g}} 
\end{equation} 
where $\int_{\bar{C}}$ is defined as in \eqref{intcC}--\eqref{epsorder1} but with $\veps$ replaced by $-\veps$ (a derivation of this formula can be found in Appendix~\ref{app:norm2}). 

\subsection{Completeness}
\label{subsec:Completeness} 
It is interesting to know if the states obtained in Proposition~\ref{Prop:sJack}  provide a complete orthogonal basis in the Fock space $\cF$. 
The following results due to van Elburg and Schoutens  \cite{ES} suggests an answer to this questions. 

(Recall that $\cH^{\nu,2}$ is the CFT Hamiltonian in \eqref{cHnu2}, and \eqref{rsdef} {\em ff.} for the definition of the integers $r,s$.) 
 
\begin{lemma}
\label{lem:ES}
Let $\eta^{N,M}_{Q,\vn}$ be a set of linearly independent states in $\Db$ labeled by $N,M\in\mathbb{N}_0$, $Q\in\mathbb{Z}$ and $\vn\in\mathbb{N}_0^{N+M}$ such that 
\begin{equation} 
\label{nrestr2} 
s-r \geq Q\geq 1-r,\quad n_1\geq \cdots \geq n_N\geq M+\chi_{Q>0},\quad n_{N+1}\geq \cdots \geq n_{N+M}\geq 0, 
\end{equation} 
with $\chi_{Q>0}=1$ for $Q>0$ and $0$ otherwise, and which obey 
\begin{equation}
\label{Heta=Eeta}
\cH^{\nu,2}\eta^{N,M}_{Q,\vn} = E^{N,M}_{2,Q,\vn} \eta^{N,M}_{Q,\vn}
\end{equation} 
with $E^{N,M}_{2,Q,\vn}$ in \eqref{E2NMnun}. Then this set of states is a complete basis in $\cF$ if and only if the following functional identity holds true, 
\begin{equation} 
\label{ESidentity} 
 \sum_{Q\in\mathbb{Z}} q^{\frac{Q^2}{2rs}} \frac1{\prod_{n=1}^\infty(1-q^n)} =  
 \sum_{Q=1-s}^{r-s}  \sum_{N,M=0}^\infty  \frac{q^{\frac{1}{2rs}Q^2+\frac{r}{2s}N^2 + \frac{s}{2r}M^2 +\frac1sQN-\frac1rQM+\chi_{Q>0}N}}{\prod_{n=1}^N(1-q^n)\prod_{m=1}^M(1-q^m)}
\end{equation}
for $q\in\mathbb{C}$ such that $|q|<1$.

\end{lemma} 

(See Appendix~\ref{subsubsec:ES} for details on how Lemma~\ref{lem:ES} follows from results in \cite{ES}. 
We apply this result below to $\eta^{N,M}_{Q,\vn}=\psi^{N,M}_{Q,\vn}$, and we use it also in Appendix~\ref{sec:cHnu3} for other states.)

The identity in \eqref{ESidentity} is of Rogers-Ramanujan type. 
We checked this remarkable identity extensively using MAPLE and thus are convinced that it is true.  
However, we do not know its mathematical status. 
The reason for why we did not invest more effort in this questions is that we cannot rule out the possibility that the states $\psi^{N,M}_{Q,\vn}$ in Proposition~\ref{Prop:sJack}, with $Q$ and $\vn$ restricted as in \eqref{nrestr2}, are linearly dependent. For this reason, Lemma~\ref{lem:ES} and Proposition~\ref{Prop:sJack} do not imply that these states are a complete orthogonal basis even if the identity in \eqref{ESidentity} is proved.  
In particular,  there exist states with degenerate eigenvalues $E^{N,M}_{k,Q,\vn}$, $k=1,2,3$, and such states could be linearly dependent (this degeneracy problem is discussed in more detail in Appendix~\ref{sec:cHnu3}). 
However, we do not believe that this happens for the following reasons. As mentioned, there exists a family of commuting differential operators $D^{(k)}_{N,M}$ of the form  
\begin{equation} 
D^{(k)}_{N,M}(\vx,\vy;g) = \sum_{j=1}^N\left(-\ii \frac{\partial}{\partial x_j} \right)^{k-1} +(-g)^{k-2}  \sum_{k=1}^N\left(-\ii \frac{\partial}{\partial y_k} \right)^{k-1} + \mbox{ lower order terms}
\end{equation} 
for $k=2,3,\ldots$ and equal to $H_{N,M}$ in \eqref{HNM} and $D_{N,M}$ in \eqref{PNM} for $k=3$ and $2$, respectively. 
Moreover, the functions in \eqref{Psin} {\em ff.} are common eigenfunctions of all these differential operators with known eigenvalues. 
We believe that all these differential operators have collective field representations by self-adjoint commuting operators $\cH^{\nu,k}$ on $\cF$, and these allow to generalize the result in Theorem~\ref{Thm:Main} to $k=4,5,\ldots$. 
In fact, a collective field representation $\cH^{\nu,4}$ of the differential operators $D^{(4)}_{N,0}$ was constructed in \cite{EPSS} (such operator is also given in \cite{ELCFT}, Remark~4 in Section~4), and in Ref.~\cite{NS} such operators are constructed for all $k$ and $M=0$ (in a framework different from ours). 
We believe that these operators are collective field representations of the deformed CS operators $D^{(k)}_{N,M}$ as well. 
If so, then the states in Proposition~\ref{Prop:sJack} are common eigenstates of all these operators $\cH^{\nu,k}$, and this would not leave any room for different such states to not be orthogonal. 
This still would not rule out the possibility that some of these states are zero. 
However, from the explicit formulas for the Hilbert space norms of these states in \eqref{psinorm}--\eqref{Pi}, we think this is unlikely. 

We thus {\bf conjecture} that {\em the states in \eqref{psiNMnQ1}--\eqref{q0}, with $N,M\in\mathbb{N}_0$, $Q\in\mathbb{Z}$ and $\mathbb{N}_0^{N,M}$ such that \eqref{nrestr2} holds true, provide a complete orthogonal basis in $\cF $.} 
 
 It is interesting to note that, according to \eqref{nQuantumNrs} and \eqref{nrestr2}, not all super Jack polynomials are needed to generate the full CFT Hilbert space $\cF$. A prominent example not needed are the super Jack polynomial $P^{N,M}_{\vzero}$ labeled by the empty partition $\vzero$. It would be interesting to understand this over-completeness of the super Jack polynomials. 

\subsection{Proof of Proposition~\ref{Prop:sJack}}
\label{subsec:ProofThm:sJack}
We start with a result which links the product of vertex operators $\Phi^{N,M}_{\nu}$ in \eqref{PhinuNM} with deformed Newton sums $p^{N,M}_{\vlam}$ in \eqref{spNMg}--\eqref{sJack}. 
For that we use the usual bijection between partitions $\vlam=(\lambda_1,\lambda_2,\ldots)$ and integer vectors $\vm=(m_1,m_2,\ldots)$, $m_n\in\mathbb{N}_0$ and only finitely many $m_n$ non-zero,  given by $\vlam =\langle 1^{m_1},2^{m_2},\ldots\rangle$, for example,  the partition $\vlam=(5,4,4,2,2,2,1,0) = \langle 1^1,2^3,4^2,5^1\rangle$ corresponds to the integer vector $\vm$ with the following non-zero components,  $m_1=m_5=1$, $m_2=3$, $m_4=2$. This allows to label the boson vectors $\eta^b_{Q,\vm}$ in \eqref{etab} by partitions $\vlam$. We find it convenient to use the notation 
\begin{equation} 
\label{vlamQ}
|Q,\vlam\rangle \equiv \eta^b_{Q,\vm}\quad (\vlam=\langle 1^{m_1},2^{m_2},\ldots\rangle)
\end{equation} 
to emphasize this. 
We also use the shorthand notation $Z_{\vlam} \equiv \prod_{n=1}^\infty m_n! n^{m_n}$, and $\ell(\vlam)$ is the number of non-zero parts $\lambda_j$ of the partition $\vlam=(\lambda_1,\lambda_2,\ldots)$.

\begin{lemma} For all $Q\in\mathbb{Z}$, 
\label{lem:useful} 
\begin{multline} 
\label{vlamstates}
\xxa\Phi^{N,M}_\nu(\vx,\vy)\xxe|Q\rangle = \ee^{-\ii q_0(|\vx|-\nu^{-2}|\vy|)} \exp\left(\nu \sum_{n>0} \frac{\cc_{-n}}{n} p_n^{N,M}(\vz^{-1},\vw^{-1};\nu^2) \right) |Q_1\rangle = \\ 
\ee^{-\ii q_0(|\vx|-\nu^{-2}|\vy|)}\sum_{\vlam} Z_{\vlam}^{-1/2}\nu^{\ell(\vlam)}p^{N,M}_{\vlam}(\vz^{-1},\vw^{-1};\nu^2) |Q_1,\vlam\rangle
\end{multline} 
with $q_0$ in \eqref{q0}, $\vz^{-1}\equiv(\ee^{-\ii x_1},\ldots,\ee^{-\ii x_N})$, $\vw^{-1}\equiv (\ee^{-\ii y_1},\ldots,\ee^{-\ii y_M})$, 
\begin{equation}
\label{Q1}  
Q_1= Q+N\nu/\nu_0-M/(\nu\nu_0), 
\end{equation} 
$|Q,\vlam\rangle$ in \eqref{vlamQ}, and the sum over all partitions $\vlam=(\lambda_1,\lambda_2,\ldots)$. 
\end{lemma} 

(The proof are straightforward computations using the definitions and \eqref{eQRw1}.) 

To show that the state  in \eqref{psiNMnQ1} is well-defined, we compute it using Lemma~\ref{lem:useful}. 
We find
\begin{equation} 
\label{psivlam}
\psi^{N,M}_{Q,\vn} =  \sum_{\vmu}   |Q_1,\vmu\rangle Z_{\vmu}^{-1/2}\nu^{\ell(\vmu)} (p^{N,M}_{\vmu},P^{N,M}_{\vlam}) 
\end{equation} 
with $Q_1$ in \eqref{Q1}, the sum over all partitions $\vmu$ satisfying $|\vmu|=|\vlam|$,  and 
\begin{equation} 
\label{pPprod}
 (p^{N,M}_{\vmu},P^{N,M}_{\vlam}) \equiv  \int_C d^Nx\, d^My\,   p^{N,M}_{\vmu}(\vz^{-1},\vw^{-1};\nu^2) \Delta^{N,M}(\vz,\vw;\nu^2) P^{N,M}_{\vlam}(\vz,\vw;\nu^2) , 
\end{equation} 
\begin{equation} 
\label{Delta}
\Delta^{N,M}(\vz,\vw;g)\equiv \frac{\prod_{j\neq j'}^N\left(1-z_j/z_{j'}\right)^g\prod_{k\neq k'}^M\left(1-w_k/w_{k'}\right)^{1/g}  }{\prod_{j=1}^N\prod_{k=1}^M (1-z_j/w_k)(1-w_k/z_j)} 
\end{equation} 
 (the interested reader can find details of this computation in Appendix~\ref{app:norm1}). 
Since the integrand is absolutely bounded on the integration domain, it is clear that the integrals in \eqref{pPprod} are well-defined. 
Since there are only finitely many partitions $\vmu$ such that $|\vmu|=|\vlam|$, this proves that the integrals in \eqref{psiNMnQ1} define states in $\Db$. 

One sees by inspection that the integrand  in \eqref{pPprod} is analytic in the variables $\vz$ and $\vw$ in the region
\begin{equation} 
\label{Eq:restr11}
\begin{split}
-\pi<\Re(x_j)<\pi\quad (1\leq j\leq N),\quad -\pi<\Re(y_k)<\pi\quad (1\leq k\leq M),\\
0>\Im(x_1)>\cdots\Im(x_N)>\Im(y_1)>\cdots>\Im(y_M),\qquad\qquad
\end{split}
\end{equation} 
and thus the integral in \eqref{pPprod} remains unchanged if we deform the integration contours so that 
\begin{equation} 
\label{epsorder2}
0<\epsilon_1 =\cdots=\epsilon_N< \epsilon_{N+1}=\cdots=\epsilon_{N+M} . 
\end{equation} 
Moreover, it follows from the definitions in \eqref{intcC}--\eqref{cCeps} that 
\begin{subequations} 
\begin{align} 
\frac{\partial}{\partial\epsilon_1} (p^{N,M}_{\vmu},P^{N,M}_{\vlam}) = -\ii 
\int_{\cC_0} d^Nx\,d^My\, \left\{ \left( \sum_{j=1}^N\frac{\partial}{\partial x_j}\right)  \Delta^{N,M}(\vz,\vw;\nu^2) f(\vz,\vw) \right\} ,
\label{BT11} \\
\frac{\partial}{\partial\epsilon_{N+1}} (p^{N,M}_{\vmu},P^{N,M}_{\vlam}) = -\ii 
\int_{\cC_0} d^Nx\,d^My\,     \left\{ \left( \sum_{k=1}^M\frac{\partial}{\partial y_k}\right)  \Delta^{N,M}(\vz,\vw;\nu^2) f(\vz,\vw) \right\} 
\label{BT22}
\end{align} 
\end{subequations} 
where $f(\vz,\vw)$ is short for $p^{N,M}_{\vmu}(\vz^{-1},\vw^{-1};\nu^2)P^{N,M}_{\vlam}(\vz,\vw;\nu^2)$, and $\cC_0$ is to indicate that $\veps$ is as in \eqref{epsorder2} (the interchange of integration and differentiation is justified by the analyticity of the integrand). 
We now use that the function in \eqref{Delta} is proportional to
\begin{equation}
\label{nonfactor} 
\begin{split} 
\frac{\prod_{j< j'}^N\left(1-z_j/z_{j'}\right)^{2g}\prod_{k <k'}^M\left(1-w_k/w_{k'}\right)^{2/g}  }{\prod_{j=1}^N\prod_{k=1}^M (1-z_j/w_k)^2}(z_1\cdots z_N)^M(w_1\cdots w_M)^{-N} \times \\ 
\exp\left( -\ii g\sum_{j=1}^N(N+1-2j)x_j -\ii(1/g)\sum_{k=1}^M(M+1-2k)y_k) \right) , 
\end{split} 
\end{equation} 
which implies that the integrals in \eqref{BT11} and \eqref{BT22} are zero (to see this for \eqref{BT11}, note that, if $F(\vz)$ is analytic and symmetric in $\vz=(\ee^{\ii x_1},\ldots,\ee^{\ii x_N})$, then 
\begin{multline} 
\int_{-\pi-\ii\epsilon_1}^{\pi+\ii\epsilon_1} dx_1
\cdots \int_{-\pi-\ii\epsilon_1}^{\pi+\ii\epsilon_1} dx_{N}\left\{\left( \sum_{j=1}^N\frac{\partial}{\partial x_j} \right) \ee^{-\ii g\sum_{j=1}^N(N+1-2j)x_j}F(\vz) \right\} = \\
\int_{-\pi-\ii\epsilon_1}^{\pi+\ii\epsilon_1} dx_1
\cdots \int_{-\pi-\ii\epsilon_1}^{\pi+\ii\epsilon_1} dx_{N}\left\{\left( \sum_{j=1}^N\frac{\partial}{\partial x_j} \right) \ee^{-\ii g\sum_{j=1}^{\lfloor N/2\rfloor}(N+1-2j)(x_j-x_{N+1-j})}F(\vz) \right\} = 
0 
\end{multline} 
since the non-trivial boundary terms coming from the partial derivatives with respect to $x_j$ and $x_{N+1-j}$ add up to zero, and similarly for \eqref{BT22}). 
This completes the proof that the integrals in \eqref{pPprod}, and thus also the states in \eqref{psiNMnQ1}, are independent of $\veps$. 

We turn to the eigenstate properties. 
We use  \eqref{cH1sOmega} and \eqref{cH1nurel} to compute 
\begin{multline} 
\cH^{\nu,1}\psi^{N,M}_{Q,\vn} =  \int_C d^Nx\,d^My\, \Psi^{N,M}_{q_0,\vn}(\vx,\vy;\nu^2) 
\bigl( [\cH^{\nu,1},\Phi^{N,M}_\nu(\vx,\vy)] +  \\ \Phi^{N,M}_\nu(\vx,\vy) \cH^{\nu,1}\bigr)|Q\rangle =E^{N,M}_{1,Q} \psi^{N,M}_{Q,\vn} 
\end{multline} 
with $E^{N,M}_{1,Q}$ in \eqref{E1NMnun}. This proves the result for the charge operator $\cH^{\nu,1}$. 

Similarly, using  \eqref{cH2sOmega} and \eqref{cH2nurel}, performing a partial integration, and using that $\Psi^{N,M}_{q_0,\vn}$ is an eigenstate of $D_{N,M}$ with eigenvalue $E^{(2)}_{\vlam}$ given in \eqref{E2lam},   
\begin{multline} 
\cH^{\nu,2}\psi^{N,M}_{Q,\vn} =  \int_C d^Nx\,d^My\, \Psi^{N,M}_{q_0,\vn}(\vx,\vy;\nu^2)
\bigl( -D_{N,M}(\vx,\vy) + \half(\nu_0Q)^2  \bigr) \Phi^{N,M}_\nu(\vx,\vy)|Q\rangle = \\
BT +  \int_C d^Nx\,d^My\, \left\{ \bigl( D_{N,M}(\vx,\vy) + \half(\nu_0Q)^2  \bigr) \Psi^{N,M}_{q_0,\vn}(\vx,\vy;\nu^2)\right\}  \Phi^{N,M}_\nu(\vx,\vy)|Q\rangle= \\ 
BT + \left(E^{(2)}_{\vlam} + \half(\nu_0 Q)^2 \right) \psi^{N,M}_{Q,\vn} = BT +E^{N,M}_{2,Q,\vn} \psi^{N,M}_{Q,\vn}
\end{multline} 
with $E^{N,M}_{2,Q,\vn}$ in \eqref{E2NMnun} and the following boundary term coming from the partial integration, 
\begin{equation} 
BT \equiv  \ii \int_C d^Nx\,d^My\, \left\{ \left(\sum_{j=1}^N \frac{\partial}{\partial x_j} + \sum_{k=1}^M \frac{\partial}{\partial y_k}\right)  \Psi^{N,M}_{q_0,\vn}(\vx,\vy;\nu^2)\Phi^{N,M}_\nu(\vx,\vy)\right\} |Q\rangle. 
\end{equation}  
Inserting \eqref{psivlam}--\eqref{pPprod} and using that the boundary terms in \eqref{BT11} and \eqref{BT22} are zero shows that $BT=0$,  which completes the proof that $\psi^{N,M}_{Q,\vn}$ is an eigenstate of $\cH^{\nu,2}$ with eigenvalue given in \eqref{E2NMnun}. 

The computation which proves the corresponding result for $\cH^{\nu,3}$ was already given in \eqref{cHnu3comp}; the boundary terms arising in the partial integrations can be shown to be zero by similar arguments as used above to show that $BT=0$. \QED

\section{Super Jack polynomials from CFT states}
\label{sec:Math}

We shortly discuss one applications of Theorem~\ref{Thm:Main}  to the mathematical theory of super Jack polynomials. 

We present a mapping from common eigenstates $\eta$ of the CFT operators $\cH^{\nu,k}$, $k=1,2,3$,  to eigenfunctions of deformed CS model. 
The key result below is a simple consequence of Theorem~\ref{Thm:Main}  and a straightforward extension of a result in \cite{CL}. 

(Recall the definitions in \eqref{spNMg}--\eqref{Psi01}, \eqref{etab} {\em ff.}, \eqref{OmQ} and \eqref{PhinuNM}.)

\begin{cor}
\label{cor:A}
Let $\eta\in \Db$ be a common eigenstate of the operators $\cH^{\nu,k}$ in \eqref{cHnu1}--\eqref{cHnu3} with corresponding eigenvalues $E_k$ for $k=1,2,3$. Then 
\begin{equation} 
f_{\eta}(\vx,\vy)\equiv \langle Q|\Phi^{N,M}_\nu(\vx,\vy)^\dag |\eta\rangle  
\end{equation} 
for 
\begin{equation} 
\label{Qcond} 
Q = \nu\nu_0^{-1}E_1 - \nu\nu_0^{-1}N + (\nu\nu_0)^{-1}M 
\end{equation} 
gives a function of the form 
\begin{equation} 
\label{feta1} 
f_\eta(\vx,\vy) = \ee^{\ii q_0(|\vx|-|\vy|/g)}\Psi_0^{N,M}(\vx,\vy;g) P^{N,M}_\eta(\vz,\vw;g) 
\end{equation} 
with $P_{\eta}^{N,M}\in\Lambda_{N,M,g}$, $q_0$ in \eqref{q0} and $g=\nu^2$. 
Moreover, $f_\eta(\vx,\vy)$ is a common eigenfunction of the differential operators $H_{N,M}$ in \eqref{HNM} and $D_{N,M}$ in \eqref{PNM} with corresponding eigenvalues 
\begin{subequations} 
\begin{equation} 
\label{E1eta} 
E_{\eta}^{(2)} = E_2 -\frac12(\nu_0 Q)^2 
\end{equation} 
and
\begin{equation} 
\label{E2eta} 
E_{\eta}^{(3)} = E_3-\frac1{12}(\nu^2-\nu^{-2})M - \frac{\nu(\nu_0Q)^3}3+\frac{\nu^3\nu_0Q}{12}, 
\end{equation} 
\end{subequations} 
respectively. 
\end{cor}   

(A proof is given in Appendix~\ref{app:ProofcorA}.) 

As an application of this result, one can take for $\eta$ the states in Proposition~\ref{Prop:sJack} but for different parameters $N',M',Q'$: 
\begin{equation} 
f^{N,M,N',M'}_{Q',\vn}(\vx,\vy) =  \langle Q|\Phi^{N,M}_\nu(\vx,\vy)^\dag |\psi^{N',M'}_{Q',\vn}\rangle 
\end{equation}
with
\begin{equation} 
Q=Q'-\nu\nu_0^{-1}(N-N')+(\nu\nu_0)^{-1}(M-M'). 
\end{equation} 
This provides a map from eigenfunctions of the dCS Hamiltonian with particle numbers $N',M'$ to eigenfunctions of a dCS Hamiltonian with other particle numbers $N,M$. 
By straightforward computations one finds that, for $\eta=\psi^{N',M'}_{Q',\vn}$, $P_{\eta}^{N,M}=\hat\Pi^{N',M'}_{N,M}P^{N',M'}_{\vlam}$ with an integral operators $\hat\Pi^{N',M'}_{N,M}$ defined on polynomials $P\in\Lambda_{N',M',g}$ as follows, 
\begin{equation} 
\label{hPi} 
(\hat\Pi^{N',M'}_{N,M} P)(\vz,\vw;g) \equiv \int_C d^{N'}x'\, d^{M'}y'\, \Pi^{N',M'}_{N,M}(\vz,\vw,(\vz')^{-1},(\vw')^{-1};g) P(\vz',\vw';g) , 
\end{equation} 
\begin{equation}
\label{Pi1} 
\Pi^{N',M'}_{N,M}(\vz,\vw,\vz',\vw';g) \equiv \frac{\prod_{j=1}^N\prod_{k'=1}^{M'}(1-z_jw'_{k'})\prod_{j'=1}^{N'}\prod_{k=1}^M(1-z'_{j'}w_k)}{\prod_{j=1}^N\prod_{j'=1}^{N'}(1-z_j z_{j'}')^{g}\prod_{k=1}^M\prod_{k'=1}^{M'}(1-w_kw_{k'}')^{1/g}} 
\end{equation} 
(this is obtained by computations similar to ones in Appendices~\ref{app:norm2} and \ref{app:norm1}). 
This provides a map between super Jack polynomials with different variable numbers $(N',M')$ and $(N,M)$, in generalization of results in \cite{SV2} (the proof of this will be given elsewhere\footnote{Work in in collaboration with Martin Halln\"as (in preparation).}). 

We mention that a similar relation between super Jack polynomials with different variable numbers was previously found in \cite{HL} using a different method. 
In fact, we checked that the results in \cite{HL} are obtained by applying the map in Corollary \ref{cor:A} to the orthogonalized anyon states $\tilde\psi^{N,M}_{Q}(\vn)$ constructed in Appendix~\ref{sec:cHnu3}. 
Thus the results in this paper provide a quantum field theory interpretation of results in \cite{HL}. 
It is worth mentioning that this clarifies some puzzling aspect of the results in \cite{HL}. 
In particular, the generalized commutator relations of the Fourier modes of the anyon operators (see \eqref{CARgen} and \eqref{CARgen2}) explain why certain states constructed in \cite{HL}, and which should not exist,  are indeed zero. 
It is worth noting that the results in \cite{HL} where not only for the (deformed) CS model, but for a large class of quantum Calogero-Moser-Sutherland models. 
It would be interesting to establish results like in this paper for other such systems. 

\newsection{Concluding remarks} 
\label{sec:Con} 
The FQHE is observed in certain 2D electron systems where the Hall conductance has plateaus at values which are fractional multiples of the natural constant $e^2/h$. 
The fractions thus measured have a striking accuracy (see e.g.\ Fig.~1 in \cite{TSG}), and a convincing theoretical explanation of this effect can therefore not be based  on approximation methods.
Physicists have met this challenge by identifying integrable structures in models that can explain the FQHE.  
We hope that the result in this paper can contribute to making further progress in this direction. 
As one specific possibility, we mention recent work proposing a occupations-number-like picture for FQHE states based on Jack polynomials \cite{BH}. 
In that work, Jack polynomials for {\em negative} rational values of the parameter $\alpha=1/g$ \cite{Stanley} are used. 
For such values of $\alpha$, the interpretation of the corresponding CS model as quantum-many body system is not possible, and the corresponding CFT is not unitary. 
Our results suggest that the deformed CS model can provide further interesting fractional quantum Hall states. 
It is tempting to speculate that the latter states for {\em positive} rational $\alpha$ could be an alternative to the states proposed in \cite{BH}, and that this would provide a link to Wen's theory \cite{Wen}. 
We hope that our result will inspire future work in this direction. 
 
Wen's theory \cite{Wen} has been very successful to explain many aspects of the FQHE. As discussed in \cite{ES}, an orthogonalized anyon basis should be very useful to compute quantities of interest for physicists  which previously were not accessible. We hope that our result in Proposition~\ref{Prop:sJack} will help to establish the theory of super Jack polynomials as a useful tool in such computations. 

Our results also motivate future research on the mathematical theory of special functions. 
For example, the formula for the Hilbert space norm of the super Jack states in \eqref{psinorm} provides a natural candidate for an inner product of the super Jack polynomials, and it would be interesting to study this product. 
It also would be interesting to generalize the result in this paper to the elliptic CS model, in generalization of results in \cite{L}. 

\subsection*{Acknowledgments} 
We would like to thank Eddy Ardonne, Alan Carey, Hans Hansson, Martin Halln\"as, Douglas Lundholm, Per Moosavi, Didina Serban and Richard Szabo for useful discussions. 
We are grateful to Martin Halln\"as for valuable suggestions on the first version of this paper. 
This work was supported by the G\"oran Gustafsson Foundation. F.A. acknowledges support from Olle Eriksson Foundation for Materials Engineering (No. VT-2015-0001).

\appendix

\newsection{Computation details and proofs} 
\label{app:Proofs} 
Some of the results used in the main text are technical or can be proved by simple adaptation of results available in the literature. In this appendix we collect some of these proofs for the convenience of the reader. 

\subsection{Apropos Section~\ref{sec:dCS}}
\label{app:eigenvals} 

\subsubsection{Eigenvalues for empty partition}
\label{app:vlam=0} 
We give details on how the constants $E^{(2,3)}_{\vzero}$ in \eqref{E2lam}--\eqref{E3lam} are obtained. 

It is obvious that $\Psi_0^{N,M}(\vx,\vy;g)$ in \eqref{Psi0} is invariant if all variables $x_j$ and $y_k$ are shifted by the same constant. Thus $D_{N,M}\Psi_0^{N,M}=0$, which implies $E^{(2)}_{\vzero}=0$. 

The constant $E^{(3)}_{\vzero}$ in \eqref{E3lam} can be obtained from Proposition~2.1 in \cite{ELell} as the special case $q=0$,  $\cN=N+M$, $m_J=1$ for $1\leq J\leq N$, and $m_J=-1/\lambda$ for $N+1\leq J\leq N+M$ (note that $g$ here corresponds to $\lambda$ in \cite{ELell}).  

\subsubsection{Eigenvalue identities} 
\label{app:vlam_vs_vn}
We explain how \eqref{E1n}--\eqref{E2n} are obtained from \eqref{E2lam}--\eqref{E3lam}. 

We first consider the special case $q_0=0$ when the function in \eqref{Psin} is identical to the one in \eqref{Psilambda}. 
In this case the eigenvalues in \eqref{E2lam}--\eqref{E3lam}  and \eqref{E1n}--\eqref{E2n} are the same, as is seen by straightforward computations using the following relations between a partition $\vmu$ and its conjugate $\vmu'$, 
\begin{equation} 
\sum_j\mu_j=\sum_j\mu_j',\quad \sum_j\mu_j^2 = \sum_j(2j-1)\mu_j',\quad \sum_j(2j-1)\mu_j=\sum_j(\mu_j')^2. 
\end{equation} 
The generalization of this to non-zero $q_0$ follows by straightforward computations. 

\subsection{Apropos Section~\ref{sec:sJack}}
\label{app:sJack} 

\subsubsection{Derivation of \eqref{psinorm}--\eqref{Pi}}
\label{app:norm2} 
To find the Hilbert space norm of the state $\psi^{N,M}_{Q,\vn}$ in \eqref{psiNMnQ1} we compute the scalar product of this state with itself:
\begin{multline} 
\label{psinorm1} 
||\psi^{N,M}_{Q,\vn}||^2 = \int_{\bar{C}} d^Nx\,d^My\,\int_C d^Nx'\, d^My'\, P_{\vlam}(\vz^{-1},\vw^{-1};\nu^2) P_{\vlam}(\vz',\vw';\nu^2) \times \\
\ee^{-\ii q_0(|\vx|-|\vx'|-\nu^{-2}(|\vy|-|\vy'|))}\Psi_0^{N,M}(-\vx,-\vy;\nu^2)\Psi^{N,M}_{0}(\vx',\vy';\nu^2) \times \\
  \langle Q|\Phi^{N,M}_\nu(\vx,\vy)^\dag \Phi^{N,M}_\nu(\vx',\vy')|Q\rangle .
\end{multline} 
We compute the function in the third line in \eqref{psinorm1} for $Q=0$ using $\phi_\nu(x)^\dag=\phi_{-\nu}(x)$ and \eqref{phinuprod}, and we find that it is equal to $\Psi_0^{N,M}(\vx,\vy;\nu^2)\Psi^{N,M}_{0}(-\vx',-\vy';\nu^2)$ times 
\begin{multline}
\label{kernelfunction} 
\frac{\prod_{j=1}^M\prod_{k=1}^M  \left(\ee^{\ii(x'_{j}-y_k)/2}-\ee^{\ii(y_k-x'_{j} )/2}\right)  \left(\ee^{\ii(y'_{k}-x_j)/2}-\ee^{\ii(x_j-y'_{k} )/2}\right) }{\prod_{j,j'=1}^N\left(\ee^{\ii(x'_{j'}-x_j)/2}-\ee^{\ii(x_j-x'_{j'} )/2}\right)^{\nu^2}\prod_{k,k'=1}^M  \left(\ee^{\ii(y'_{k'}-y_k)/2}-\ee^{\ii(y_k-y'_{k'} )/2}\right)^{1/\nu^{2}}} = \\
\ee^{\ii(N\nu^2-M)(|\vx|-|\vx'|-\nu^{-2}(|\vy|-|\vy'|))/2}\frac{\prod_{j=1}^N\prod_{k=1}^M(1-w_k/z_j')(1-z_j/w_k')}{\prod_{j<j'}^N(1-z_j/z_{j'}')^{\nu^2}\prod_{k<k'}^M(1-w_k/w_{k'}')^{1/\nu^2}}. 
\end{multline}
Recalling $q_0$ in \eqref{q0} one sees that the exponential factor here exactly cancels the one in \eqref{psinorm1}.
Inserting \eqref{PsiDelta}  we obtain \eqref{psinorm}--\eqref{Pi}. 
It is straightforward to generalize this computation to non-zero $Q$, and one finds that the result does not depend on $Q$. 

\subsubsection{Derivation of \eqref{psivlam}--\eqref{Delta}}
\label{app:norm1} 
Note that
\begin{equation} 
\label{PsiDelta} 
\Delta^{N,M}(\vz,\vw;g)  = \Psi_0^{N,M}(\vx,\vy;g) \Psi_0^{N,M}(-\vx,-\vy;g) 
\end{equation} 
with $\Psi_0^{N,M}$ in \eqref{Psi01} and $z_j=\ee^{\ii x_j}$, $w_k=\ee^{\ii y_k}$. 

Inserting \eqref{nphiNM}  and Lemma~\ref{lem:useful} in \eqref{psiNMnQ1} and using \eqref{PsiDelta} we obtain  
\begin{equation} 
\psi^{N,M}_{Q,\vn} =  \int_C d^Nx\, d^My\,   \Delta^{N,M}(\vz,\vw;g) P^{N,M}_{\vlam}(\vz,\vw;\nu^2) 
\sum_{\vmu} Z_{\vmu}^{1/2} \nu^{\ell(\vmu)}p^{N,M}_{\vmu}(\vz^{-1},\vw^{-1};\nu^2) |Q_1,\vmu\rangle . 
\end{equation} 
Interchanging summation and integration we obtain \eqref{psivlam}--\eqref{pPprod}. 
This interchange is justified by interpreting $\psi^{N,M}_{Q,\vn} $ as en element in the dual of $\Db$, i.e.\ we only consider $\langle\eta|\psi^{N,M}_{Q,\vn} \rangle$ with $\eta\in\Db$ (this is clear since, with the scalar product with $\eta$ in place, the $\vmu$-sum has at most finitely many non-zero terms).

We now show that the integral defined in \eqref{pPprod} satisfies 
\begin{equation} 
\label{pPprod11}
(|\vlam|-|\vmu|) (p^{N,M}_{\vmu},P^{N,M}_{\vlam}) =0, 
\end{equation} 
which proves that the sum in \eqref{psivlam} can be restricted to the partitions $\vmu$ satisfying $|\vmu|=|\vlam|$. 
Indeed,  using $D_{N,M}$ in \eqref{PNM} and  $f(\vz,\vw) = p^{N,M}_{\vmu}(\vz^{-1},\vw^{-1};\nu^2)P^{N,M}_{\vlam}(\vz,\vw;\nu^2)$, the following identity holds true
\begin{equation} 
\left\{ D_{N,M}(\vx,\vy) \Delta^{N,M}(\vz,\vw;\nu^2)f(\vz,\vw) \right\}  = \\
 \left(|\vlam|-|\vmu| \right)\Delta^{N,M}(\vz,\vw;\nu^2)f(\vz,\vw)
\end{equation} 
(this is obtained by inserted the definition of $\Delta^{N,M}$, the Leibniz rule, and \eqref{DPlamP}).  
Applying to this the integration in \eqref{intcC}, recalling \eqref{pPprod},  and using that the integrals in \eqref{BT11} and \eqref{BT22} are zero,  one obtains \eqref{pPprod11}. 

\subsubsection{Completeness}
\label{subsubsec:ES}
We obtained Lemma~\ref{lem:ES} from a result in \cite{ES}, Appendix~A using the following dictionary between the notation in \cite{ES} and ours:   
\begin{equation}
\begin{split} 
M^{ES}=N,\quad N^{ES}=M,\quad Q^{ES}=-Q,\quad r^{ES}=r,\quad s^{ES}=s ,\\
 m_j^{ES}=n_{N+1-j}+M \quad (1\leq j\leq N),\quad n_k^{ES} = n_{N+M+1-k}\quad (1\leq k\leq M) . 
\end{split} 
\end{equation} 
We checked that, with these identifications, the eigenvalues of $\cH^{\nu,3}$ in \eqref{E3NMnun} agree with the results reported in \cite{ES}, Section~V.A. 
   
Lemma~\eqref{lem:ES} can be proved by computing the partition function $Z\equiv \mathrm{Tr}_{\cF}(\exp(-\beta \cH^{\nu,2}))$ for the operators $\cH^{\nu,2}$ in \eqref{cHnu2} in tho different ways: First, using the boson states $\eta^b_{Q,\vm}$ in \eqref{etab}, which are a complete orthonormal basis of eigenstates of $\cH^{\nu,2}$ with corresponding eigenvalues $\frac{1}{2rs}Q^2 + \sum_{n>0} nm_n$;  $Z$ computed in this way is equal to the l.h.s.\ in \eqref{ESidentity} with $q=\ee^{-\beta}$. Second, using the states $\eta^{N,M}_{Q,\vn}$ and \eqref{Heta=Eeta} with $E^{N,M}_{2,Q,\vn}$ in \eqref{E2NMnun} and \eqref{pjdef}; straightforward computations show that $Z$ then becomes equal to the r.h.s.\ in \eqref{ESidentity} if the labels are constrained as in \eqref{nrestr2}.  This proves Lemma~\ref{lem:ES}. \QED

\subsection{Apropos Section~\ref{sec:Math}}

\subsubsection{Proof of Corollary~\ref{cor:A}}
\label{app:ProofcorA}
For $k=1,2,3$, compute $\langle \eta|\cH^{\nu,k} \Phi^{N,M}_\nu(\vx,\vy)|Q\rangle $ in two different ways: 
First, by taking the adjoint of  Theorem~\ref{Thm:Main}  and \eqref{cHnusOmega} and using that $\cH^{\nu,k}$ is self-adjoint, and second,  by acting with $\cH^{\nu,k}$ on the state $\eta$ using that it is an eigenstate with eigenvalue $E_k$. This gives  
\begin{subequations} 
\begin{align} 
\left( (N-M\nu^{-2})+ \frac{\nu_0Q}{\nu} \right)f_\eta(\vx,\vy)  = E_1f_\eta(\vx,\vy),  \\
\left( D_{N,M}(\vx,\vy) + \frac12(\nu_0 Q)^2 \right)f_\eta(\vx,\vy)  = E_2f_\eta(\vx,\vy), \\
\left( H_{N,M}(\vx,\vy;\nu^2) +\frac1{12}(\nu^2-\nu^{-2}) + \frac{\nu(\nu_0Q)^3}3-\frac{\nu^3\nu_0Q}{12} \right)f_\eta(\vx,\vy)  = E_3f_\eta(\vx,\vy)
\end{align} 
\end{subequations} 
for $k=1,2,3$, respectively. The first of these equations show that $f_{\eta}(\vx,\vy)$ can be non-zero only if the condition in \eqref{Qcond} is fulfilled. 
The other two equations prove that $f_{\eta}(\vx,\vy)$ is a common eigenfunction of the differential operators $H_{N,M}$ and $D_{N,M}$ and determines the corresponding eigenvalues. 

Inserting \eqref{nphiNM} and \eqref{vlamstates} gives  
\begin{equation} 
f_\eta(\vx,\vy) = \ee^{\ii q_0(|\vx|-\nu^{-2}|\vy|)}\Psi_0^{N,M}(\vx,\vy;g) \sum_{\vlam}p^{N,M}_{\vlam}(\vz,\vw;\nu^2) \langle Q_1,\vlam|\eta\rangle 
\end{equation} 
with $q_0$ in \eqref{q0} and $Q_1=Q+N\nu/\nu_0-M/(\nu\nu_0)$. Note that the sum has at most a finite number of non-zero terms according to Lemma 4.3.
This proves that $f_\eta(\vx,\vy)$ is as in \eqref{feta1} with $P^{N,M}_\eta\in\Lambda_{N,M,\nu^2}$.

\newsection{Orthogonalized anyon basis} 
\label{sec:cHnu3} 
One initial motivation for the project leading to this paper was to work out the construction of anyon basis suggested in \cite{ES} but for all particles numbers $(N,M)$, in generalization of previous results \cite{CL}. 
In this appendix we present this generalization since, as we believe, is a useful complement to the result in Proposition~\ref{Prop:sJack}: it is less elegant but, at the same time, more elementary in that it does not rely on mathematical results about the super Jack polynomials. Moreover, in combination with Corollary~\ref{cor:A}, it gives explicit formulas for the super Jack polynomials, and this provides an alternative proof of results previously obtained by other methods in \cite{HL}. 

\subsection{Anyon basis}
\label{subsec:eigenstates} 
Recall \eqref{rsdef} and that $\Db$ is the set of {\em finite} linear combinations of boson states as in \eqref{etab}. 

For generic $\nu_0$ and $\nu\in\nu_0\mathbb{Z}$, the vertex operator $\phi_\nu(x)$ in \eqref{phinu} is not $2\pi$-periodic in the variable $x$ due the presence of the factors $\exp(-\ii\nu \cQ x/2)$. 
However, the modified vertex operators 
\begin{equation}
\label{cphinu}
\check\phi_\nu(x)\equiv   \ee^{\ii\nu \cQ x/2}  \phi_\nu(x)  \ee^{\ii\nu \cQ x/2} 
\end{equation} 
are  $2\pi$-periodic and can be  Fourier transformed as follows, 
\begin{equation} 
\label{hatphinu1} 
\hat\phi_\nu(n) \equiv \int_{-\pi - \ii \epsilon}^{\pi - \ii \epsilon}  \check\phi_\nu(x) \ee^{\ii nx} dx, \quad (\epsilon>0, n\in\mathbb{Z}). 
\end{equation} 
The Fourier modes $\hat\phi_\nu(n)$ are well-defined operators on $\cF$:
\begin{lemma} 
\label{lemma:hatphin} 
\label{lemmaphinu} 
For all $n\in\mathbb{Z}$ and $\nu\in\nu_0\mathbb{Z}$, $\hat\phi_\nu(n)$ defined in \eqref{hatphinu1} is independent of $\epsilon$ and a well-defined operator mapping $\Db$ to $\Db$. 
In particular, for all states $|Q\rangle = R^Q|0\rangle$,   
\begin{equation} 
\label{hatphinu0}
\hat\phi_\nu(0)|Q\rangle = |Q+\nu/\nu_0 \rangle \quad (Q\in\mathbb{Z}). 
\end{equation} 
\end{lemma} 
(See Section~\ref{subsubsec:ProofLem} for proof.) 

Set $\nu=r\nu_0$ so that $1/\nu=s\nu_0$. Then Lemma~\ref{lemma:hatphin} implies that, for arbitrary $N,M\in\mathbb{N}_0$, $Q\in\mathbb{Z}$ and $\vn=(n_1,\ldots,n_{N+M})\in\mathbb{Z}^{N+M}$, the state
\begin{equation} 
\label{etaNMnQ}
\eta^{N,M}_{Q}(\vn)\equiv \hat\phi_\nu(n_1)\cdots \hat\phi_\nu(n_N)\hat\phi_{-1/\nu}(n_{N+1})\cdots \hat\phi_{-1/\nu}(n_{N+M})|Q\rangle 
\end{equation} 
is well-defined and in $\Db$. 
Due to the highest weight condition in \eqref{HWcond}, this state is non-zero under certain restrictions on the integer vectors $\vn$: 

\begin{lemma} 
\label{lem:HW} 
The state $\eta_{Q}^{N,M}(\vn)$ defined in \eqref{etaNMnQ} is in $\Db$ and non-zero only if all the following conditions on the integer vector $\vn$ are fulfilled, 
\begin{equation}
\label{nrestr}
n_j+n_{j+1}+\cdots + n_{N+M}\geq 0 \quad \forall j=1,2,\ldots, N+M. 
\end{equation}
\end{lemma} 
(See Section~\ref{subsubsec:HW} for proof). 

We observe that the states in \eqref{etaNMnQ} can be computed as 
\begin{multline} 
\label{etaNMnQ1} 
\eta_{Q}^{N,M}(\vn) = \int_{-\pi-\ii\epsilon_1}^{\pi-\ii\epsilon_1} dx_1\, \ee^{\ii n^+_1 x\pdag_1} \cdots  \int_{-\pi-\ii\epsilon_N}^{\pi-\ii\epsilon_N} dx_N\, \ee^{\ii n^+_N x\pdag_N}  
\int_{-\pi-\ii\epsilon_{N+1}}^{\pi-\ii\epsilon_{N+1}} dy_1\, \ee^{\ii n^+_{N+1} y\pdag_1} \cdots  \times \\
\int_{-\pi-\ii\epsilon_{N+M}}^{\pi-\ii\epsilon_{N+M}} dy_M\, \ee^{\ii n^+_{N+M} y\pdag_M} \phi_\nu(x_1)\cdots  \phi_\nu(x_N)\phi_{-1/\nu}(y_1)\cdots  \phi_{-1/\nu}(y_M)|Q\rangle 
\end{multline} 
with $\epsilon_j$ constrained as in \eqref{epsorder1} and $n^+_j$ are as in \eqref{pjdef} (see Section~\ref{subsubsec:ProofEqs}). 
We can use the abbreviation in \eqref{PhinuNM} to write \eqref{etaNMnQ1} as
\begin{equation} 
\label{etaNMnQ2} 
\eta_{Q}^{N,M}(\vn) = \int_{-\pi-\ii\epsilon_1}^{\pi-\ii\epsilon_1} dx_1\cdots \, 
\int_{-\pi-\ii\epsilon_{N+M}}^{\pi-\ii\epsilon_{N+M}}  dx_{N+M} \, \ee^{\ii\vn^+\cdot\vx} \, \Phi_\nu^{N,M}(\vx,\vy)|Q\rangle 
\end{equation} 
where $\vn^+\cdot \vx$ is short for $\sum_{j=1}^{N+M} n_j^+x\pdag_j$ with $x_j=y_{j-N}$ for $j>N$. 
It is therefore clear that one can use Theorem~\ref{Thm:Main}  and \eqref{cHnusOmega} to compute the action of the operators $\cH^{\nu,1}$, $\cH^{\nu,2}$ and $\cH^{\nu,3}$ in \eqref{cHnu1}--\eqref{cHnu3} on the states $\eta_{Q}^{N,M}(\vn)$. 
Straightforward computations lead to the following results. 

\begin{cor} 
\label{Cor:cHketa} 
For arbitrary $N,M\in\mathbb{N}_0$, let $\eta_{Q}^{N,M}(\vn) $ be as in \eqref{etaNMnQ} with $\vn\in \mathbb{Z}^{N+M}$ and $Q\in\mathbb{Z}$. Then    
\begin{subequations} 
\begin{align}
\label{cH1res22}
 \cH^{\nu,1}\eta_{Q}^{N,M}(\vn)  =&E^{N,M}_{1,Q}\eta_{Q}^{N,M}(\vn) ,\\
 \label{cH2res22} 
 \cH^{\nu,2} \eta_{Q}^{N,M}(\vn) =& E^{N,M}_{2,Q,\vn}\eta_{Q}^{N,M}(\vn) ,\\
 \label{cH3res22} 
 \cH^{\nu,3} \eta_{Q}^{N,M}(\vn)  =& E^{N,M}_{3,Q,\vn} \eta_{Q}^{N,M}(\vn)  - 2\sum_{j<k}^{N+M} \gamma_{jk} \sum_{\mu=1}^\infty\mu\,  \eta_{Q}^{N,M}(\vn + \mu[\ve_j-\ve_k])
\end{align} 
\end{subequations} 
with $E^{N,M}_{1,Q}$, $E^{N,M}_{2,Q,\vn}$, $E^{N,M}_{3,Q,\vn}$ in \eqref{EkNMnun}, $n^+_j$ in \eqref{pjdef}, $(\ve_j)_\ell=\delta_{j,\ell}$, and $\gamma_{jk}$ in \eqref{gammajk}. 
\end{cor} 

(The proof is given in Appendix~\ref{subsubsec:proofCor}.)

It is important to note that the states in \eqref{etaNMnQ} are not linearly independent: the operators $\hat\phi_{\nu}(n)$ and $\hat\phi_{-1/\nu}(n)$ obey generalized commutator relations \cite{BLS} (we give these relations in Appendix~\ref{subsubsec:CARgen}) and, as suggested by results discussed in Section~\ref{subsec:Completeness}, one can restrict the quantum numbers $\vn$ and $Q$ by the conditions in \eqref{nrestr2} {\em ff}. Note that \eqref{nrestr} is automatically fulfilled if \eqref{nrestr2} holds true, but this former restrictions still is important since it shows that the sum on the r.h.s.\ in \eqref{cH3res22}�   at most has a finite number of non-zero terms and thus is well-defined. 

\subsection{Orthogonalization}
\label{subsec:orth} 
Corollary~\ref{Cor:cHketa} shows that the states  $\eta^{N,M}_{Q}(\vn)$ are special in that they are common eigenfunctions of the operators  $\cH^{\nu,1}$ and $\cH^{\nu,2}$. Thus one expects that these states, with $Q$ and $\vn$ restricted as in \eqref{nrestr2}, provide a useful basis in applications like, for example, computations of quantities of interest for physicists in Wen's effective theory of the FQHE \cite{Wen}. 
However, this basis has a serious drawback: it is not orthogonal. 
As suggested in \cite{ES}, one can improve this basis by finding linear combinations of these states $\eta^{N,M}_{Q}(\vn)$ which also are eigenstates of $\cH^{\nu,3}$. 
Examples of this orthogonalization were given in \cite{ES} for $(N,M)=(1,1)$ and in \cite{CL} for $(N,M)=(N,0)$, $N$ arbitrary,  and $(0,M)$, $M$ arbitrary.  
We now extend this construction to all $(N,M)$ using \eqref{cH3res22}. 
As we will see, this extension is non-trivial, i.e., there is an additional complication as compared to the special cases $(N,M)=(N,0)$, $(0,M)$, and $(1,1)$.

We now explain the construction of these improved states. 
For that we note that $\cH^{\nu,3}$ acts on the states $\eta^{N,M}_{Q}(\vn)$ in \eqref{etaNMnQ} like a triangular matrix if one uses the partial ordering $\preceq$ between integer vectors $\vn,\vm$ in $\mathbb{Z}^{N+M}$ defined as follows, 
\begin{equation} 
\label{porder} 
\vm\preceq\vn : \Leftrightarrow m_j+\cdots+m_{N+M}\leq n_j+\cdots n_{N+M}\quad \forall j=1,2,\ldots,N+M. 
\end{equation} 
Indeed, it is clear from \eqref{cH3res22} that, if one acts repeatedly with  $\cH^{\nu,3}$ on $\eta^{N,M}_{Q}(\vn)$, one always obtains a linear combination of states $\eta^{N,M}_{Q}(\vm)$ with 
\begin{equation} 
\label{vm}
 \vm=\vn+\sum_{j<k}^{N+M}\mu_{jk} (\ve_j-\ve_k),\quad \mu_{jk}\in\mathbb{N}_0\quad \forall j<k, 
\end{equation} 
and $\vm\preceq\vn$ for all such $\vm$. This suggests that it is possible to construct eigenstates $\tilde\psi^{N,M}_{Q}(\vn)$ of $\cH^{\nu,3}$ as linear superpositions of these states  $\eta^{N,M}_{Q}(\vm)$  with coefficients $u_{\vn}(\vm)$ which can be computed by diagonalizing a triangular matrix:
\begin{equation} 
\label{psiNMnQ2} 
\tilde\psi^{N,M}_{Q}(\vn) = \eta^{N,M}_{Q}(\vn)  + \sum_{\vm\prec \vn} u_{\vn}(\vm)\eta^{N,M}_{Q}(\vm) . 
\end{equation} 
Inserting this ansatz into the eigenvalue equation $(\cH^{\nu,3}-E)\tilde\psi^{N,M}_{Q}(\vn)=0$ and using \eqref{cH3res22} gives 
\begin{multline} 
\label{algorithm} 
\sum_{\vm\preceq \vn} u_{\vn}(\vm)\left( (E^{N,M}_{3,Q,\vm}-E)\eta^{N,M}_{Q}(\vm)  - 2\sum_{j<k}^{N+M} \gamma_{jk} \sum_{\mu=1}^\infty\mu \eta_{Q}^{N,M}(\vm + \mu[\ve_j-\ve_k]) \right) = \\
\sum_{\vm\preceq \vn} \eta^{N,M}_{Q}(\vm)\left( (E^{N,M}_{3,Q,\vm}-E)u_{\vn}(\vm) - 2\sum_{j<k}^{N+M} \gamma_{jk} \sum_{\mu=1}^\infty\mu u_{\vn}(\vm - \mu[\ve_j-\ve_k]) \right) = 0
\end{multline} 
with $u_{\vn}(\vn)=1$. This shows that the eigenvalue equation is fulfilled provided that  $E=E^{N,M}_{3,Q,\vm}$, $u_{\vn}(\vm)=0$ for $\vm\succ\vn$,  and, using the definition
\begin{equation} 
\label{bnm} 
b_{\vn}(\vm)\equiv E^{N,M}_{3,Q,\vm} - E^{N,M}_{3,Q,\vn}, 
\end{equation} 
\begin{equation} 
\label{unm}
 u_{\vn}(\vm) = \frac{2}{b_{\vn}(\vm)}\sum_{j<k}^{N+M} \gamma_{jk} \sum_{\mu=1}^\infty\mu u_{\vn}(\vm - \mu[\ve_j-\ve_k]) 
\end{equation} 
for $\vm\prec \vn$ provided $b_{\vn}(\vm)\neq 0$. It is important to note that, due to Lemma~\ref{lem:HW}, $\eta^{N,M}_{Q}(\vm)$ is non-zero only for finitely many $\vm$ as in \eqref{vm}, and thus there are only finitely many cases $\vm\prec\vn$ to be considered (for other $\vm$ one can set $u_{\vn}(\vm)=0$). 
 
Thus eigenstates of $\cH^{\nu,3}$ can be constructed in this manner provided the following {\bf non-degeneracy condition} is fulfilled: {\em The integer vector $\vn$ is such that, for all $\vm$ of the form \eqref{vm} fulfilling the conditions $\sum_{k=j}^{N+M}m_j\geq 0$ for all $j=1,2,\ldots,N+M$, the eigenvalue differences $b_{\vn}(\vm)$ in \eqref{bnm} are non-zero.} 

At this point the restriction in \eqref{nrestr2} becomes important: inserting \eqref{E3NMnun} into \eqref{bnm} one finds for the special cases $(N,M)=(N,0)$, $(0,M)$ and $(1,1)$ mentioned above 
\begin{equation} 
\label{bnvm}
b_n(\vm)  = \begin{cases}
 \sum_{j=1}^N\mu_j^2 + \sum^N_{j<j'}\mu_{jj'}(n_j-n_{j'}+\nu^2(j'-j)) & (M=0) \\
 -\nu^2\sum_{k=N+1}^{N+M}\mu_{j}^2 -\nu^2 \sum^M_{k<k'}\mu_{kk'}(n_j-n_{j'}+\nu^{-2}(k'-k)) & (N=0) \\ 
\mu_{12}^2 +2\mu_{12}(n_1-1) + \nu^2\mu_{12} (2n_2-\mu_{12})+\mu_{12}(1 +\nu^2) & (N=M=1)  
 \end{cases} 
\end{equation} 
with $\mu_j\equiv \sum_{k>j}\mu_{jk}-\sum_{k<j}\mu_{kj}$. This proves that, in these special cases, the non-degeneracy condition is always fulfilled provided \eqref{nrestr2} holds true: 
 Since $n_j>n_{j'}$ for $1\leq j<j'\leq N$ and $n_k>n_{k'}$ for $N+1\leq k<k'\leq N+M$, $b_{\vn}(\vm)$ is manifestly positive and manifestly negative in the cases $(N,M)=(N,0)$ and $(0,M)$, respectively. 
Moreover, for the case $(N,M)=(1,1)$, positivity of $b_{\vn}(\vm)$ follows from $m_2=n_2-\mu_{12}\geq 0$ and  $n_1\geq 1$. 
Unfortunately, for general $(N,M)$, we cannot rule out $b_{\vn}(\vm)=0$ by \eqref{nrestr2} in such a simple manner.  
However, we stress that we expect that degeneracies are an exception: generically, the algorithm above should work (but we cannot prove this in general). 

We summarize the construction above as follows (recall the definition of $\preceq$ in \eqref{porder}). 

\begin{prop}
\label{Prop:cHkpsi}
Let $Q\in\mathbb{Z}$ and $\vn\in\mathbb{N}_0^{N+M}$ such that the no-degeneracy condition above is fulfilled, e.g., $(N,M)=(N,0)$, $(0,M)$ or $(1,1)$, and $Q$ and $\vn$ satisfy the conditions in \eqref{nrestr2}. 
Then a unique state $\tilde\psi^{N,M}_{Q}(\vn)\in\Db$ is given by \eqref{psiNMnQ2} with coefficients $u_{\vn}(\vm)$ determined by the following finite recursion procedure: $u_{\vn}(\vm)=0$ if $\vm\succ\vn$, $u_{\vn}(\vn)=1$, $u_{\vn}(\vm)$ is given by \eqref{unm}--\eqref{bnm} if $\vm$ is as in \eqref{vm} such that $\sum_{k=j}^{N+M}m_k\geq 0$ for $j=1,2,\ldots,N+M$, and $u_{\vn}(\vm)=0$ otherwise. 
Moreover, provided this state $\tilde\psi^{N,M}_{Q}(\vn)$  is non-zero, it is a common eigenstate of the operators $\cH^{\nu,1}$, $\cH^{\nu,2}$ and $\cH^{\nu,3}$ in \eqref{cHnu1}--\eqref{cHnu3} with corresponding eigenvalues given in \eqref{EkNMnun}.
\end{prop}

(The proof is given in the paragraphs containing \eqref{porder}--\eqref{bnm}, except for the common eigenfunction property. The latter is a simple consequence of results stated in Corollary~\ref{Cor:cHketa}.) 

As emphasized in Proposition~\ref{Prop:cHkpsi}, it is possible that a state $\tilde\psi^{N,M}_{Q}(\vn)$ constructed with this algorithm is zero. As an example we mention a case $(N,M)=(2,0)$ where the algorithm gives the following   state $\tilde\psi(\vn)\equiv \tilde\psi^{2,0}_{\nu,0}(\vn)$ for $\vn=(1,2)$ (note that this $\vn$ does not satisfy the condition in \eqref{nrestr2}), 
\begin{equation} 
\tilde\psi(1,2) = \tilde\psi(1,2) + (g-1)\tilde\psi(2,1)+\half g(g-1)\tilde\psi(3,0) , 
\end{equation} 
but this state can be shown to be zero by using the generalized commutator relation in \eqref{CARgen}. 
However, we found that this happens only for states with integer vectors $\vn$ {\em not} satisfying the condition in \eqref{nrestr2}. 

\begin{remark} We note that it is possible to modify the algorithm described in the Proposition~\ref{Prop:cHkpsi} as follows (in the following we denote the set of all integer vectors $\vn$ satisfying the condition in \eqref{nrestr2} as $S$):  in cases where a  state $\eta^{N,M}_{Q}(\vm)$ in \eqref{algorithm} appears where $\vm\notin S$, rewrite this state as a linear combination of states $\eta^{N,M}_{Q}(\vk)$ with $\vk\in S$ before solving for $u_{\vn}(\vm)$ (this rewriting can be done using \eqref{CARgen}--\eqref{CARgen2}). Doing this, the non-degeneracy condition $b_{\vn}(\vm)\neq 0$ would only have to hold true for integer vectors $\vm\in S$ appearing in this modified algorithm, and this is probably fulfilled for many more integer vectors $\vn\in S$.
\end{remark} 

\subsection{Details and proofs}
\label{app:ProofsSec:cHnu3} 

\subsubsection{Proof of Lemma~\ref{lemma:hatphin}}
\label{subsubsec:ProofLem} 
One can compute $\hat\phi_\nu(n)$ by inserting 
\begin{equation} 
\label{phinu1} 
\begin{split} 
 \ee^{\ii\nu \cQ x/2} \phi_\nu(x)  \ee^{-\ii\nu \cQ x/2} =  \prod_{k>0}\exp{\left(\nu  \frac{\cc_{-k}}{k}\ee^{-\ii kx}\right)}\exp{\left(-\nu\frac{\cc_k}{k}\ee^{\ii kx}\right)} R^{\nu/\nu_0}  
\end{split} 
\end{equation} 
in the definition  \eqref{cphinu}--\eqref{hatphinu1}, expanding the exponentials in Taylor series, and performing the $x$-integration (note that \eqref{phinu1} follows from \eqref{phinu}). 
This computation can be made mathematically precise by interpreting the operators as quadratic forms on $\Db$ (a collection of mathematical results needed for that can be found in \cite{LM}, Appendix~C). 
It thus is clear that $\hat\phi_\nu(n)$, for arbitrary $\nu\in\nu_0\mathbb{Z}$, is a linear combination of terms 
\begin{equation} 
\label{hatphinun2}
\cc_{-\ell_1}\cc_{-\ell_2}\cdots \cc_{-\ell_L}\delta_{\ell_1+\ell_2+\cdots+\ell_L,n}R^{\nu/\nu_0}  
\end{equation} 
with non-zero integers $\ell_j$ such that $ \ell_1\geq \ell_2\geq \cdots \geq \ell_L$ (note that the $\ell_j$ can be negative), and only a finite number of these terms give a non-zero result when acting on a state in $\Db$; see Appendix~C.1 in \cite{CL} for precise formulas. This shows that $\hat \phi_\nu(n)$ is well-defined on $\Db$. It is easy to prove that operators as in \eqref{hatphinun2} map $\Db$ to $\Db$, and thus the same is true for $\hat\phi_\nu(n)$.   

In particular, using \eqref{phinu1} and the definition of $|Q\rangle$,
\begin{equation} 
\hat\phi_\nu(0)|Q\rangle = \int_{-\pi-\ii\epsilon}^{\pi-\ii\epsilon} dx\,  \prod_{k>0}\exp{\left(\nu  \frac{\cc_{-k}}{k}\ee^{-\ii kx}\right)}|Q+\nu/\nu_0\rangle = |Q+\nu/\nu_0\rangle
\end{equation} 
since $\cc_{k}|Q\rangle=0$ for all $k>0$.

\begin{remark}
A alternative proof with more explicit formulas can be found in \cite{CL}, Appendix~C.1 (it is easy to see that the different regularization used in \cite{CL} does not affect the result in any way).   
\end{remark} 

\subsubsection{Proof of Lemma~\ref{lem:HW}} 
\label{subsubsec:HW}
For $m\in\mathbb{N}_0$ and $Q\in\mathbb{Z}$,  let $\mathcal{D}_{m,Q}$ be the vector space of all finite linear combinations of states 
\begin{equation} 
\cc_{-m_1}\cc_{-m_2}\cdots \cc_{-m_M}|Q\rangle 
\end{equation} 
with $M\in\mathbb{N}_0$ and $m_j\in\mathbb{N}$ such that $m_1+\cdots+m_M=m$. This is obviously a subset of $\Db$, and $\Db$ is equal to the union of all these spaces. 
Using repeatedly $\cc_{-\ell}\cc_{-m_j}=\cc_{-m_j}\cc_{-\ell}-\ell\delta_{m_j+\ell,0}$ and $\cc_{\ell}|Q\rangle=0$ for $\ell>0$, one sees that, for all $\eta\in\mathcal{D}_{m,Q}$, $\cc_{-\ell}\eta\in\mathcal{D}_{m+\ell,Q}$ for $m+\ell\geq 0$ and $\cc_{-\ell}\eta=0$ otherwise. 
Thus the operators in \eqref{hatphinun2} map $\mathcal{D}_{m,Q}$ to $\mathcal{D}_{m+n,Q}$ if $m+n\geq 0$ and to $0$ otherwise and, since $\hat\phi_\nu(n)$ is a finite linear combination of such operators (see Section~\ref{subsubsec:ProofLem}), the same is true for $\hat\phi_\nu(n)$. 
Thus the states in \eqref{etaNMnQ}�   are either $0$ or in $\mathcal{D}_{|n|,Q}$ with $|n|=\sum_{j=1}^{N+M}n_j$, and they are non-zero only if the conditions in \eqref{nrestr} are fulfilled. \QED

\begin{remark}
Lemma~\ref{lem:HW} is a generalization of Proposition~3 in \cite{CL}. 
\end{remark} 

\subsubsection{Proof of Equations \eqref{etaNMnQ1}--\eqref{etaNMnQ2}}
\label{subsubsec:ProofEqs} 
We note the identity 
\begin{equation} 
\label{eQRw1}
\ee^{\ii c\cQ} R^Q = \ee^{\ii \nu_0 c Q} R^Q \ee^{\ii c\cQ} \quad (Q\in\mathbb{Z}, c\in\mathbb{C})
\end{equation} 
which follows from the second relation in \eqref{CCR}. 
Using this,  and since $\langle\eta,\Phi^{N,M}_\nu(\vx,\vy)|0\rangle $ is an analytic function for all $\eta\in\Db$ in the region defined \eqref{Eq:restr}, it is clear that the r.h.s.\ in \eqref{etaNMnQ} can be written as an integral on the r.h.s.\ in \eqref{etaNMnQ1} with $\epsilon_j$ as in \eqref{epsorder1} and computable parameters $n^+_j$. To compute the these parameters we use that, for all $K\in\mathbb{N}$, $\nu_j\in\nu_0\mathbb{Z}$, and suitable complex $x_j$, 
\begin{multline} 
\ee^{\ii\nu_1 \cQ x_1/2}  \phi_{\nu_1}(x_1)  \ee^{\ii\nu_1 \cQ x_1/2} \cdots  \ee^{\ii\nu_K \cQ x_K/2}  \phi_{\nu_K}(x_K)  \ee^{\ii\nu_K \cQ x_K/2} =   \\ 
\ee^{ \ii \sum_{j=1}^K \left(\frac12 \nu_j^2  + \sum_{k=j+1}^K  \nu_j\nu_k \right)x_j} 
\phi_{\nu_1}(x_1)\cdots  \phi_{\nu_K}(x_K)  \ee^{ \ii \sum_{j=1}^K \nu_j x_j \cQ  } , 
\end{multline} 
which is easily proved by induction using \eqref{eQRw1}. Setting $K=N+M$, 
\begin{equation}
\label{nujdef}  
\nu_j = \begin{cases} \nu & (1\leq j\leq N) \\ -1/\nu & (N+1\leq j\leq N+M) ,\end{cases}  
\end{equation} 
we find  
\begin{equation} 
n^+_j = n_j + \nu_0\nu_j Q+ \frac12\nu_j^2 + \sum_{k=j+1}^{N+M}\nu_j\nu_k.
\end{equation} 
This is equal to $n_j^+$ in \eqref{pjdef}, as can be shown by straightforward computations. \QED

\subsubsection{Proof of Corollary~\ref{Cor:cHketa}} 
\label{subsubsec:proofCor} 
To prove the results for $\cH^{\nu,3}$ we use Theorem~\ref{Thm:Main}  and \eqref{cHnusOmega} to find  
\begin{multline} 
\cH^{\nu,3} \Phi_\nu^{N,M}(\vx,\vy)|Q\rangle = \left( \frac{1}{12}(\nu^2+\nu^{-2})M +  \frac{\nu(\nu_0Q)^3}3-\frac{\nu^3\nu_0Q}{12} \right) \Phi_\nu^{N,M}(\vx,\vy)|Q\rangle + \\ 
H_{N,M}(\vx,\vy;\nu^2) \Phi_\nu^{N,M}(\vx,\vy)|Q\rangle  .
\end{multline} 
Inserting this in \eqref{etaNMnQ2} we obtain 
\begin{multline} 
\cH^{\nu,3}\eta_{Q}^{N,M}(\vn) = \left( \frac{1}{12}(\nu^2+\nu^{-2})M +  \frac{\nu(\nu_0Q)^3}3-\frac{\nu^3\nu_0Q}{12} \right) \eta_{Q}^{N,M}(\vn)+\\ 
%\int_{-\pi-\ii\epsilon_1}^{\pi-\ii\epsilon_1} dx_1\, \ee^{\ii n^+_1 x_1} \cdots  
%\int_{-\pi-\ii\epsilon_{N+M}}^{\pi-\ii\epsilon_{N+M}} dy_M\, \ee^{\ii n^+_{N+M} y_M} 
 \int_{-\pi-\ii\epsilon_1}^{\pi-\ii\epsilon_1} dx_1\cdots \, 
\int_{-\pi-\ii\epsilon_{N+M}}^{\pi-\ii\epsilon_{N+M}}  dx_{N+M} \, \ee^{\ii\vn^+\cdot\vx} \, H_{N,M}(\vx,\vy;\nu^2)\Phi_\nu^{N,M}(\vx,\vy)|Q\rangle . 
\end{multline} 
To compute the latter integral we write the differential operator in \eqref{HNM} as 
\begin{equation} 
H_{N,M}(\vx,\vy;\nu^2) = -\sum_{j=1}^K \frac{\nu}{\nu_j}\frac{\partial^2}{\partial x_j^2} - 2\sum_{1\leq j<k\leq K}\gamma_{jk}\sum_{\mu=1}^\infty \mu\ee^{\ii\mu(x_j-x_k)}
\end{equation} 
using notations introduced in \eqref{DefProof} and \eqref{gammajk} and the representation of $\half \sin^{-2}(\half x)$ in \eqref{Vdef} (note that this representation is adequate in the integration domain due to our assumption in \eqref{epsorder1}). 
We thus obtain \eqref{cH3res22}  with $E^{N,M}_{3,Q,\vn}$  in \eqref{E3NMnun}. 
  
The proofs of the results for  $\cH^{\nu,1}$ and $\cH^{\nu,2}$ are similar but simpler and thus omitted. \QED

\subsubsection{Generalized commutator relations}
\label{subsubsec:CARgen}
One can show that the following generalized commutator relations hold true on $\Db$, 
\begin{equation} 
\label{CARgen}
 \sum_{\ell=0}^\infty \binom{-\mu^2}{\ell}(-1)^\ell\Bigl( \hat\phi_{\mu}(n+\ell)\hat\phi_{\mu}(m-\ell) -   \hat\phi_{\mu}(m+\ell)\hat\phi_{\mu}(n-\ell) \Bigr)=0
\end{equation} 
for $\mu=\nu$ and $-1/\nu$, and 
\begin{equation} 
\label{CARgen2} 
\hat\phi_\nu(n)\hat\phi_{-1/\nu}(m)+ \hat\phi_{-1/\nu}(m+1)\hat\phi_\nu(n-1) = \hat\phi_{\nu-1/\nu}(n+m). 
\end{equation} 
By using repeatedly these relations and Lemma~\ref{lem:HW}, any state $\eta^{N,M}_{Q}(\vn)$ in \eqref{etaNMnQ} can be written as a {\em finite} linear combination of such states with integer vectors $\vn$ satisfying $n_1\geq \cdots \geq n_N\geq M$ and $n_{N+1} \geq \cdots\geq n_{N+M}\geq 0$. The number of independent such states is further reduced by \eqref{hatphinu0}, which implies
\begin{equation} 
\label{MK} 
\eta^{N,M}_{Q}(\vn) = \eta^{N,K}_{Q-(M-K)s}(\vn) \; \mbox{ if } \; n_{N+K+1}=\cdots = n_{N+M}=0 
\end{equation} 
for all integers $K$ in the range $0\leq K\leq M$ (recall that $1/(\nu_0\nu)=s$). 
To find the restriction on $Q$ so as to get a complete basis of linearly independent states is non-trivial. 
A solution of this problem was given in \cite{ES}, Appendix~A; see Lemma~\ref{lem:ES}.

\end{document}